\documentclass[superscriptaddress,twocolumn,amsmath,amssymb,prd,floatfix,nofootinbib]{revtex4}
\usepackage[dvipdfmx]{graphicx}% Include figure files
\usepackage{dcolumn}% Align table columns on decimal point
\usepackage[varg]{txfonts}
\usepackage{bm}% bold math

\usepackage{color}% draft version

\newcommand*\Laplace{\mathop{}\!\mathbin\bigtriangleup}

\begin{document}
%\begin{CJK}{UTF8}{} % Use default fonts from CJK (see below)

\title{Weakly non-Gaussian formula for the Minkowski functionals in
  general dimensions}

\author{Takahiko Matsubara} \email{tmats@post.kek.jp}
\affiliation{%
  Institute of Particle and Nuclear Studies, High Energy
  Accelerator Research Organization (KEK), Oho 1-1, Tsukuba 305-0801,
  Japan}%
\affiliation{%
  The Graduate University for Advanced Studies (SOKENDAI),
  Tsukuba, Ibaraki 305-0801, Japan}%

\author{Satoshi Kuriki} \email{kuriki@ism.ac.jp}
\affiliation{%
  Institute of Statistical Mathematics, Research Organization of
  Information and Systems, 10-3 Midoricho, Tachikawa, Tokyo 190-8562,
  Japan}%
% \affiliation{%
%   Kavli IPMU (WPI), UTIAS, The University of Tokyo, Kashiwa,
%   Chiba 277-8583, Japan}%

\date{\today}% It is always \today, today,

\begin{abstract}
  The Minkowski functionals are useful statistics to quantify the
  morphology of various random fields. They have been applied to
  numerous analyses of geometrical patterns, including various types
  of cosmic fields, morphological image processing, etc. In some
  cases, including cosmological applications, small deviations from
  the Gaussianity of the distribution are of fundamental importance.
  Analytic formulas for the expectation values of Minkowski
  functionals with small non-Gaussianity have been derived in limited
  cases to date. We generalize these previous works to derive an
  analytic expression for expectation values of Minkowski functionals
  up to second-order corrections of non-Gaussianity in a space of
  general dimensions. The derived formula has sufficient generality to
  be applied to any random fields with weak non-Gaussianity in a
  statistically homogeneous and isotropic space of any dimensions.
\end{abstract}

%\pacs{
% 98.80.-k,
% 98.65.-r,
% 98.80.Cq,
% 98.80.Es
% }% PACS, the Physics and Astronomy

\maketitle

%\end{CJK}

\section{\label{sec:Intro} Introduction}

Statistical analyses of random fields are of importance in a broad
range of research fields. For example, cosmic structures observed in
the universe, such as the temperature fluctuations of the cosmic
microwave background, the density fields in the large-scale structure,
weak lensing fields, and so forth, are considered as realizations of
random fields. Only the statistical properties of the fields can be
predicted using cosmological theories of the very early universe, such
as the theory of inflation \cite{Gut81,Sat81,Lin82,Alb82}. Although
numerous inflationary theories have been proposed to date
\cite{Mar14}, it is still not clear which theory is relevant to our
Universe. Alternatives to the theory of inflation, such as an
ekpyrotic scenario \cite{Kho01}, have also been proposed. Different
theories predict different statistical properties of the initial
density field, and thus the observational constraints against
properties of cosmic fields are crucial in discriminating between the
theories of the very early universe.

The maximal extraction of statistical information from the observed
fields is one of the most important challenges in cosmology. The most
fundamental statistic is the spatial correlation function, or its
Fourier counterpart, the power spectrum, which characterizes the
clustering strength of random fields as a function of scale
\cite{Pee80}. The two-point correlation function (or the power
spectrum) completely characterizes the statistical information of
Gaussian random fields. However, much of the information in generally
non-Gaussian random fields cannot be captured solely using this
statistic. Specifically, while simple models of single-field inflation
with a minimal kinetic term and a smooth potential predict a
negligible level of non-Gaussianities \cite{Gan94,Mal03}, various
other inflationary models can predict various types of non-negligible
non-Gaussianities \cite{Bar04,Mar14}. Therefore, it is crucial to
determine whether non-Gaussianities are contained in the initial
density field or not, and the type of non-Gaussianity if it exists.
Even if the initial density field is purely Gaussian, gravitationally
nonlinear evolution induces the non-Gaussianity in the cosmic fields,
from which one can also extract information on the evolution of the
Universe. For these reasons, non-Gaussianities, which cannot be probed
using the correlation function or the power spectrum, play important
roles in cosmology.

Straightforward statistics beyond the two-point correlation function
are the higher-order correlation functions, such as the three-point
correlation function, four-point correlation function, and so on
\cite{Pee80}. The Fourier counterparts of higher-order correlation
functions are the polyspectra, such as the bispectrum, trispectrum,
and so on \cite{Ber02}. All the statistical information of random
fields is contained in the hierarchy of these higher-order statistics.
It is relatively straightforward to theoretically predict the
polyspectra of a given model of non-Gaussian cosmic fields.
Higher-order correlation functions have many arguments because they
represent spatial correlations among many separations. Therefore, it
is challenging to measure their accurate functional forms based on
observational data.

Statistical tools for probing non-Gaussianities of random fields are
not confined to higher-order correlation functions and the
polyspectra. Among various statistical approaches, the
characterization of the morphological structures of random fields is a
unique way to probe non-Gaussianities. The Minkowski functionals
\cite{Min03,Sch93} comprise of a set of statistics that quantitatively
characterize the stochastic geometry. According to Hadwiger's theorem
\cite{Had57,Kla95}, the $d+1$ numbers of the Minkowski functionals in
$d$ spatial dimensions completely characterize the global
morphological properties that satisfy motional invariance and
additivity.

The Minkowski functionals were first introduced into cosmology by
Mecke, Buchert and Wagner \cite{MBW94} for the analysis of point sets,
such as the positions of galaxies in the Universe. Later, Schmalzing
\& Buchert \cite{SB97} consider the Minkowski functionals of excursion
sets in smoothed cosmic fields. One of the Minkowski functionals is
the Euler characteristic, or equivalently, the genus statistic of
isocontours. Prior to the introduction of the Minkowski functionals in
cosmology, the genus statistic \cite{GMD86} was applied to smoothed
cosmic fields such as the distribution of galaxies
\cite{Got87,Wei87,Got89,Par92,Vog94,Mat96,Ben01,Par05,Kim09,Jam12,Par14,Spe15,Wat17,App18,App20},
fluctuations of the cosmic microwave background
\cite{Smo94,Kog95,Col96,Par98,Par01,Col03,Col15}, weak lensing fields
\cite{Mat01,Sat03}, intergalactic medium \cite{Lee08,Hon14,Wan15}, and
the X-ray remnant of supernovae \cite{Sat19}. Applications of the
Minkowski functionals to cosmology are currently quite popular,
including the analyses of the large-scale structure
\cite{Ker97,Ker98,Sah98,Sch99,Ker01,She03,Hik03,Sha04,Hik06,Ein11,Liu20},
cosmic microwave background
\cite{Nov99,Nov00,Wu01,Pol02,Sha02,Kom03,Eri04,Nat10,Hik12,Duc13,Planck16a,Planck16b,Planck2018},
weak lensing fields
\cite{Sat01,Tar02,Kra12,Mun12,Pet13,Shi14,Mun14,Osa15,Pet15,Mun16,Shi17,Maw20,Par20},
reionization epochs
\cite{Gle06,Gle08,Fri11,McD13,Yos17,Bag18,Che19,Bag19}, and so forth.

The Minkowski functionals are also applied to other fields of
research, such as in morphological image analysis \cite{Mic01} to
describe porous media and complex fluids \cite{Mec96,Arm19}, magnetic
resonance imaging (MRI) \cite{Can09,Lar14}, the structure of human
radial peri-papillary capillaries \cite{Bar19}, mammary gland tissue
\cite{Mat07}, spinodal decomposition \cite{Mec97}, quantum motion in
billiards \cite{Kol00}, regional seismicity realizations \cite{Mak00},
microemulsions \cite{Lik98}, thin polymer films \cite{Jac98}, the
internal structure of bimetallic nanocomposites \cite{Tsu20}, the
thermodynamics of two-phase systems \cite{Ede18}, and many others.

A striking feature of the Minkowski functionals of smoothed fields is
that the shapes of the functional dependencies of the isocontour
threshold are universal for Gaussian random fields. Analytic
expressions of the Minkowski functionals for Gaussian random fields
were derived by Tomita \cite{Tom86}. Deviations from Tomita's formula
imply the non-Gaussianity of the distribution. Thus, the Minkowski
functionals are considered as probes for non-Gaussianities in cosmic
fields. Analytic expressions of the Minkowski functionals for weakly
non-Gaussian random fields in two- and three-dimensional space were
derived by one of the authors of this paper \cite{TM94,TM03}. The
expressions can be generalized to include the anisotropic effects of
redshift-space distortions of the large-scale structure
\cite{TM96,Cod13}. These formulas assume that the non-Gaussianity is
sufficiently weak and the non-Gaussian corrections are given by linear
terms of the skewness parameters. We refer to these lowest-order
corrections due to the skewness parameters as first-order corrections
of non-Gaussianity. The next-order corrections, which we call
second-order corrections of non-Gaussianity, are given by quadratic
terms of the skewness parameters and linear terms of the kurtosis
parameters. Analytic expressions of the Minkowski functionals with
second-order corrections of non-Gaussianity in a two-dimensional space
were also derived by one of the authors of this paper \cite{TM10}.
Analytic expressions of the genus statistic with second-order
corrections in two- and three-dimensional spaces were derived in
Refs.~\cite{PGP09,GPP12,Cod13}. Remarkably, formal expressions of the
genus statistic in two- and three-dimensional spaces using the
Gram-Charlier expansion to all orders are also known
\cite{PGP09,GPP12}.

The purpose of this paper is to derive analytic expressions of the
Minkowski functionals with first- and second-order corrections of
non-Gaussianity in general dimensions for the first time. In a
previous paper, Ref.~\cite{TM03}, analytic expressions of the
Minkowski functionals with first-order corrections of non-Gaussianity
in general dimensions were conjectured based on results for two- and
three-dimensions. We provide a proof of this conjecture in this paper.
In addition, we derive the second-order corrections in general
dimensions for the first time. The derived formula is given by
Eq.~(\ref{eq:4-6}) in the following, which is the main result of this
paper. The known formulas in the literature indicated above are
reproduced as special cases of the general formula.

In cosmology, the newly derived formula with second-order corrections
in three dimensions should be useful for future applications.
Mathematically, it is interesting that there exist analytic
expressions in general dimensions. Moreover, the approach used for the
derivation in this study is instructive. It is straightforward to
derive formulas for third- and higher-order corrections of
non-Gaussianity using the method presented in this work.

The authors are preparing an accompanying paper \cite{KM20}, where an
equivalent formula to this paper is mathematically derived by an
approach different from the one used in this paper. The
parameterizations of the formula in these two approaches are also
different. We have confirmed that the two expressions are actually
equivalent, and thus the derived formula is cross-checked in two
independent ways.

This paper is organized as follows. In Sec.~\ref{sec:General}, a
theory to derive non-Gaussian correction terms for a given statistic
is reviewed. In Sec.~\ref{sec:Minkowski}, the general properties and
formulas for the Minkowski functionals are reviewed. After necessary
preparations in the preceding sections, we describe the calculation of
the non-Gaussian corrections of Minkowski functionals, and present the
main result of this paper in Sec.~\ref{sec:TheFormula}. The main
conclusions are summarized in Sec.~\ref{sec:Conclusions}.
Appendix~\ref{app:LemmaA} and \ref{app:LemmaB} are devoted to the
proofs of important formulas that are used in this work.

\section{\label{sec:General} A general theory of non-Gaussian
  corrections to a mean value}

In this section, a theory to derive non-Gaussian correction terms for
a given statistic is reviewed. The same method is already described in
Ref.~\cite{TM03}.

We consider a function $F(f_\mu)$ which depends on the field value
$f(\bm{x})$ and its spatial derivatives up to the second order
\begin{equation}
    \label{eq:1-1}
    \left( f_\mu \right) =
    \left( f, f_i, f_{ij} \right)
\end{equation}
where $f_i \equiv \partial f/\partial x_i$,
$f_{ij} \equiv \partial^2f/\partial x_i \partial x_j$ ($i\leq j$), and
$x_i$ are the spatial coordinates. In $d$-dimensions, the number of
elements of the vector $f_\mu$ is $N=1 + d + d(d+1)/2=(d+1)(d+2)/2$.
We assume that the field $f$ has a zero mean, $\langle f \rangle = 0$,
which implies $\langle f_\mu \rangle = 0$. The joint probability
distribution function of the variables $f_\mu$ is denoted by
$P(\bm{f})$, and the partition function is defined by
\begin{equation}
  \label{eq:1-2}
  Z(\bm{J}) = \int_{-\infty}^\infty
  d^N\!f\, P(\bm{f})\,e^{i\bm{J}\cdot\bm{f}},
\end{equation}
where $\bm{J} = (J^\mu)$ is an $N$-dimensional vector. According to
the cumulant expansion theorem \cite{Ma85}, we have
\begin{equation}
  \label{eq:1-3}
  \ln Z(\bm{J}) =
  \sum_{n=0}^\infty \frac{i^n}{n!}
  \sum_{\mu_1=1}^N \cdots \sum_{\mu_n=1}^N
  \left\langle f_{\mu_1} \cdots f_{\mu_n} \right\rangle_\mathrm{c}
  J^{\mu_1}\cdots J^{\mu_n},
\end{equation}
where $\langle \cdots \rangle_\mathrm{c}$ denotes the cumulant.
Denoting the covariance of $\bm{f}$ as an $N\times N$ matrix $\bm{M}$
with components given as
$M_{\mu\nu} \equiv \langle f_\mu f_\nu \rangle_\mathrm{c}$, we have
\begin{multline}
  \label{eq:1-4}
  Z(\bm{J}) =
  \exp\left(
    -\frac{1}{2}\bm{f}^\mathrm{T}\bm{M}\bm{f}
  \right.
  \\
  \left.
    + \sum_{n=3}^\infty \frac{i^n}{n!}
    \sum_{\mu_1,\ldots,\mu_n}
    \left\langle f_{\mu_1} \cdots f_{\mu_n} \right\rangle_\mathrm{c}
    J^{\mu_1}\cdots J^{\mu_n}
  \right),
\end{multline}

Applying the inverse transform of Eq.~(\ref{eq:1-2}), and substituting
into the preceding equation, we obtain
\begin{align}
  P(\bm{f})
  &= \int_{-\infty}^\infty
  \frac{d^N\!J}{(2\pi)^N}
  Z(\bm{J})\, e^{-i\bm{J}\cdot\bm{f}}
  \nonumber\\
  &=
    \exp\left[
    \sum_{n=3}^\infty \frac{(-1)^n}{n!}
    \left\langle f_{\mu_1} \cdots f_{\mu_n} \right\rangle_\mathrm{c}
    \frac{\partial^n}
    {\partial f_{\mu_1}\cdots \partial f_{\mu_n}}
    \right]
    P_\mathrm{G}(\bm{f}),
  \label{eq:1-5}
\end{align}
where
\begin{align}
  P_\mathrm{G}(\bm{f})
  &=
  \int_{-\infty}^\infty \frac{d^N\!J}{(2\pi)^N}
  \exp\left(
    -i\bm{J}\cdot\bm{f}
    - \frac{1}{2} \bm{J}^\mathrm{T} \bm{M} \bm{J}
    \right)
    \nonumber\\
  &=
  \frac{1}{(2\pi)^{N/2} \sqrt{\det\bm{M}}}
  \exp\left(
    - \frac{1}{2} \bm{f}^\mathrm{T} \bm{M}^{-1} \bm{f}
    \right)
  \label{eq:1-6}
\end{align}
is the multivariate Gaussian distribution function of $\bm{f}$ with
the covariance matrix $\bm{M}$, and the repeated indices of
$\mu_1,\ldots,\mu_n$ are assumed to be summed without summation
symbols. Using the expressions of Eqs.~(\ref{eq:1-5}), the mean value
$\langle F \rangle$ of an arbitrary function $F(\bm{f})$ is formally
given by
\begin{align}
  \left\langle F \right\rangle
  &=
    \int_{-\infty}^\infty d^N\!f F(\bm{f}) P(\bm{f})
    \nonumber\\
  &=
  \left\langle
    \exp\left(
      \sum_{n=3}^\infty \frac{1}{n!}
      \left\langle f_{\mu_1} \cdots f_{\mu_n} \right\rangle_\mathrm{c}
      \frac{\partial^n}
      {\partial f_{\mu_1}\cdots \partial f_{\mu_n}}
    \right)
    F(\bm{f})
  \right\rangle_\mathrm{G},
  \label{eq:1-7}
\end{align}
where
\begin{equation}
  \label{eq:1-8}
  \left\langle \cdots \right\rangle_\mathrm{G} =
  \int_{-\infty}^\infty d^N\!f \cdots P_\mathrm{G}(\bm{f})
\end{equation}
denotes averaging over the Gaussian distribution function. Assuming
the higher-order cumulants are small, Eq.~(\ref{eq:1-7}) is formally
expanded as
\begin{multline}
  \label{eq:1-9}
  \left\langle F \right\rangle =
  \left\langle F \right\rangle_\mathrm{G} +
  \frac{1}{6}
  \left\langle f_{\mu_1}f_{\mu_2}f_{\mu_3} \right\rangle_\mathrm{c}
  \left\langle
    \frac{\partial^3 F}
    {\partial f_{\mu_1}\partial f_{\mu_2}\partial f_{\mu_3}}
  \right\rangle_\mathrm{G}
  \\
  + \frac{1}{24}
  \left\langle f_{\mu_1}f_{\mu_2}f_{\mu_3}f_{\mu_4} \right\rangle_\mathrm{c}
  \left\langle
    \frac{\partial^4 F}
    {\partial f_{\mu_1}\partial f_{\mu_2}\partial f_{\mu_3}\partial f_{\mu_4}}
  \right\rangle_\mathrm{G}
  \\
  + \frac{1}{72}
  \left\langle f_{\mu_1}f_{\mu_2}f_{\mu_3} \right\rangle_\mathrm{c}
  \left\langle f_{\mu_4}f_{\mu_5}f_{\mu_6} \right\rangle_\mathrm{c}
  \\
  \times
  \left\langle
    \frac{\partial^6 F}
    {\partial f_{\mu_1}\partial f_{\mu_2}\partial f_{\mu_3}
    \partial f_{\mu_4}\partial f_{\mu_5}\partial f_{\mu_6}}
  \right\rangle_\mathrm{G}
  + \cdots.
\end{multline}

It is useful to define dimensionless fields
\begin{equation}
  \label{eq:1-10}
  \alpha(\bm{x}) = \frac{f(\bm{x})}{\sigma_0}, \quad
  \eta_{i}(\bm{x}) = \frac{f_i(\bm{x})}{\sigma_1}, \quad
  \zeta_{ij}(\bm{x}) = \frac{f_{ij}(\bm{x})}{\sigma_2},
\end{equation}
where $\sigma_j$ are spectral parameters that are defined as
\begin{equation}
    \label{eq:1-11}
    {\sigma_j}^2 \equiv
    \left\langle f \left(-\Laplace\right)^j f\right\rangle,
\end{equation}
and $\Laplace = \partial^2/\partial x_i\partial x_i$ is the Laplacian
operator. In particular,
${\sigma_0}^2 \equiv \langle f^2 \rangle$ is the variance
of the field. We denote the set of dimensionless variables as
\begin{equation}
  \label{eq:1-12}
  \left(X_\mu\right) = \left(\alpha,\eta_i,\zeta_{ij}\right).
\end{equation}
In most of the physical applications, it is often the case that the
$m$-point cumulant of $X_\mu$ has the order of ${\sigma_0}^{m-2}$:
\begin{equation}
  \label{eq:1-13}
  \langle X_{\mu_1} \cdots X_{\mu_m} \rangle_\mathrm{c} \sim
  \mathcal{O}\left({\sigma_0}^{m-2}\right).
\end{equation}
This ordering is called hierarchical ordering. We assume this type of
ordering throughout this paper. In this case, the normalized cumulants
\begin{equation}
  \label{eq:1-14}
  C^{(m)}_{\mu_1\cdots\mu_m} \equiv
  \frac{\langle X_{\mu_1} \cdots X_{\mu_m} \rangle_\mathrm{c}}
  {{\sigma_0}^{m-2}}
\end{equation}
have order 1 in terms of $\sigma_0$.
Changing the variables from $f_\mu$ to $X_\mu$ in Eq.~(\ref{eq:1-9}),
we have a series expansion
\begin{multline}
  \label{eq:1-15}
  \left\langle F \right\rangle =
  \left\langle F \right\rangle_\mathrm{G} +
  \frac{1}{6}
  C^{(3)}_{\mu_1\mu_2\mu_3}
  \left\langle
    \frac{\partial^3 F}
    {\partial X_{\mu_1}\partial X_{\mu_2}\partial X_{\mu_3}}
  \right\rangle_\mathrm{G} \sigma_0
  \\
  +
  \left[
    \frac{1}{24}
    C^{(4)}_{\mu_1\mu_2\mu_3\mu_4}
    \left\langle
      \frac{\partial^4 F}
      {\partial X_{\mu_1}\partial X_{\mu_2}\partial X_{\mu_3}\partial X_{\mu_4}}
    \right\rangle_\mathrm{G}
    \right.
    \\
    \left.
      + \frac{1}{72}
      C^{(3)}_{\mu_1\mu_2\mu_3} C^{(3)}_{\mu_4\mu_5\mu_6}
      \left\langle
        \frac{\partial^6 F}
        {\partial X_{\mu_1}\partial X_{\mu_2}\partial X_{\mu_3}
          \partial X_{\mu_4}\partial X_{\mu_5}\partial X_{\mu_6}}
      \right\rangle_\mathrm{G}
    \right] {\sigma_0}^2
    \\
    + \mathcal{O}\left( {\sigma_0}^3 \right).
\end{multline}

\section{\label{sec:Minkowski}
Minkowski functionals
}

In this section, we define Minkowski functionals and briefly review
their properties and relations to the Euler characteristic. The
Minkowski functionals comprise of $d+1$ numbers that characterize the
morphological properties of random fields in a domain $\mathcal{D}$ of
$d$-dimensional space. For an excursion set $\mathcal{F}_\nu$, a set
of all points $\bm{x}$ with $\alpha(\bm{x}) \geq \nu$, we denote the
Minkowski functionals per unit volume as $V^{(d)}_k(\nu)$, where
$k=0,1,\ldots,d$, and their mean values as
$\bar{V}^{(d)}_k(\nu) = \langle V^{(d)}_k(\nu) \rangle$.

For $k=0$, the Minkowski functional $V^{(d)}_0$ corresponds to
the volume fraction of the excursion set,
\begin{equation}
  \label{eq:1-30}
  V^{(d)}_0(\nu)
  = \frac{1}{|\mathcal{D}|}
  \int_{\mathcal{F}_\nu} d^d\!x
   = \frac{1}{|\mathcal{D}|}
  \int_\mathcal{D} d^d\!x\,\Theta\left[\alpha(\bm{x})-\nu\right],
\end{equation}
where $|\mathcal{D}|$ is the entire volume of the domain
$\mathcal{D}$. The other Minkowski functionals with $k = 1,\ldots,d$
correspond to surface integrals of the boundary
$\partial\mathcal{F}_\nu$ of the excursion set,
\begin{equation}
  \label{eq:1-31}
  V^{(d)}_k(\nu)
  = \frac{1}{|\mathcal{D}|}
  \int_{\partial \mathcal{F}_\nu} d^{d-1}\!x\,
  v^{(d)}_k(\nu,\bm{x}),
\end{equation}
where $v^{(d)}_k(\nu,\bm{x})$ are the local Minkowski functionals
defined by
\begin{equation}
  \label{eq:1-32}
  v^{(d)}_k(\nu,\bm{x})
  = \frac{1}{\omega_k d}
    K^{(d)}_{k-1}(\nu,\bm{x}),
\end{equation}
and
\begin{equation}
  \label{eq:1-33}
  \omega_k = \frac{\pi^{k/2}}{\Gamma(k/2+1)}
\end{equation}
is the volume of the unit ball in $k$ dimensions. On the boundary
hypersurface, $\bm{x} \in \partial\mathcal{F}_\nu$,
$K^{(d)}_m(\nu,\bm{x})$ is the invariant obtained from the inverse
radii of curvature $R_1, R_2, \ldots R_{d-1}$ of the hypersurface
orientated towards the lower density regions \cite{Tom90}. That is,
\begin{equation}
  \label{eq:1-34}
  K^{(d)}_m(\nu,\bm{x})
  = \frac{1}{{}_{d-1}C_m}
  \sum_s \frac{1}{R_{s(1)}R_{s(2)}\cdots R_{s(m)}},
\end{equation}
where $\sum_s$ denotes the symmetric summation over ${}_{d-1}C_m =
(d-1)!/n!(d-m-1)!$ combinations of $m$ different components of
$(R_1,R_2,\ldots,R_{d-1})$. For example, in two-dimensional space with
$d=2$,
\begin{equation}
  \label{eq:1-35}
  K^{(2)}_0 = 1, \quad
  K^{(2)}_1 = \frac{1}{R_1},
\end{equation}
and, in three-dimensional space with $d=3$,
\begin{equation}
  \label{eq:1-36}
  K^{(3)}_0 = 1, \quad
  K^{(3)}_1 =  \frac{1}{2}
  \left(\frac{1}{R_1} + \frac{1}{R_2}\right), \quad
  K^{(3)}_2 =  \frac{1}{R_1R_2}.
\end{equation}
Substituting Eqs.~(\ref{eq:1-34}) and (\ref{eq:1-35}) into
Eqs.~(\ref{eq:1-32}) and (\ref{eq:1-31}), we obtain the formulas for
the Minkowski functionals presented in Refs.~\cite{Sch97} and
\cite{Sch98}, respectively.

The quantity
$K^{(d)}_{d-1} = {R_1}^{-1} {R_2}^{-1}\cdots {R_{d-1}}^{-1}$
corresponds to the Gauss total curvature. Due to the Gauss-Bonnet
theorem \cite{NS83}, the density of the Euler characteristic of
$\mathcal{F}_\nu$ is given by
\begin{equation}
  \label{eq:1-37}
  \chi^{(d)}(\mathcal{F}_\nu) = \frac{1}{\omega_dd}
  \frac{1}{|\mathcal{D}|}
  \int_{\partial\mathcal{F}_\nu}d^{d-1}\!x\,K^{(d)}_{d-1}(\nu,\bm{x}).
\end{equation}
Therefore, the Minkowski functional with $k=d$ corresponds to the
$d$-dimensional Euler characteristic, $V^{(d)}_d = \chi^{(d)}$. For
analytic evaluations of the Minkowski functionals of an excursion set
in random fields, Crofton's formula \cite{Cro1868,Sch97,Adl07} in
integral geometry serves as a powerful tool. This formula states that
\begin{equation}
  \label{eq:1-38}
  V^{(d)}_k(\nu)
  = \frac{\omega_d}{\omega_{d-k}\omega_k}
  \int_{\mathcal{E}^{(d)}_k} d\mu_k(E)
  \chi^{(k)}(\mathcal{F}_\nu\cap E),
\end{equation}
where $E$ is an arbitrary $k$-dimensional hypersurface and
$\chi^{(k)}$ is the density of the Euler characteristic of the
intersection $\mathcal{F}_\nu\cap E$ in $k$ dimensions. This quantity
is integrated over the space $\mathcal{E}^{(d)}_k$ of all conceivable
hypersurfaces, and the integration measure $d\mu_k(E)$ is normalized
to give $\int_{\mathcal{E}^{(d)}_k}d\mu_k(E) = 1$. Using Crofton's
formula of Eq.~(\ref{eq:1-38}), and assuming statistical isotropy and
homogeneity, the expectation values of the Minkowski functionals are
given as
\begin{equation}
  \label{eq:1-39}
  \bar{V}^{(d)}_k(\nu)
  = \frac{\omega_d}{\omega_{d-k}\omega_k}
  \left\langle \chi^{(k)}(\mathcal{F}_\nu \cap E) \right\rangle
  = \frac{\omega_d}{\omega_{d-k}\omega_k}
  \bar{V}^{(k)}_k(\nu),
\end{equation}
where $V^{(k)}_k(\nu) = \chi^{(k)}(\mathcal{F}_\nu \cap E)$ is the
$k$th-order Minkowski functional of the intersection
$\mathcal{F}_\nu \cap E$. The expectation value of $V^{(k)}_k$ does
not depend on the choice of the hypersurface $E$ due to the
statistical isotropy and homogeneity. Using this relation, the
expectation values of the Minkowski functionals for each order can be
estimated by only evaluating expectation values of the Euler
characteristic in the spaces of lower dimensions.

It is convenient to use the Morse theorem \cite{Mil63,NS83,Adl07} to
evaluate the expectation value of the Euler characteristic. The Euler
characteristic $\chi(\mathcal{F}_\nu)$ is given by an alternating sum
of the number of critical points,
\begin{equation}
  \label{eq:1-50}
  \chi(\mathcal{F}_\nu) = \sum_{m=0}^d (-1)^{d-m} C_m(\mathcal{F}_\nu),
\end{equation}
where $C_m$ is the number of critical points that satisfy
$f_i=\partial f/\partial x_i=0$ of index $m$, and the index $m$ is the
number of negative eigenvalues of the matrix
$f_{ij}=\partial^2f/\partial x_i\partial x_j$ at each critical point.

To count the number of critical points, we first consider the delta
function $\delta^d(\bm{x}-\bm{x}_\mathrm{c})$. Applying the Taylor
expansion near the critical point, we have
\begin{equation}
  \label{eq:1-51}
  f(\bm{x}) \simeq
  f(\bm{x}_\mathrm{c}) +
  \frac{1}{2} f_{ij}(\bm{x}_\mathrm{c})
  (x_i - x_{\mathrm{c}i})
  (x_j - x_{\mathrm{c}j}),
\end{equation}
up to the second order. The first-order term does not appear because a
critical point $\bm{x}_\mathrm{c}$ satisfies
$f_i(\bm{x}_\mathrm{c})=0$. Taking spatial derivatives of this
equation, we have
\begin{equation}
  \label{eq:1-52}
  \eta_i(\bm{x}) \simeq
  \frac{\sigma_2}{\sigma_1} \zeta_{ij}(\bm{x}_\mathrm{c})
  (x_j - x_{\mathrm{c}j}).
\end{equation}
Therefore, the delta function of the critical point is given by
\begin{equation}
  \label{eq:1-53}
  \delta^d(\bm{x} - \bm{x}_\mathrm{c})
  = \left(\frac{\sigma_2}{\sigma_1}\right)^d
  \delta^d(\bm{\eta})
  \left|\det\zeta\right|,
\end{equation}
near the critical point $\bm{x}_\mathrm{c}$. When the right-hand side
(rhs) is expanded to include the entire space, the left-hand side
(lhs) should be replaced by a summation of delta functions for all the
critical points. From the definition of index $m$, we have
$|\det\zeta | = (-1)^m\det\zeta$. Therefore, we have
\begin{equation}
  \label{eq:1-54}
  \sum_{\mathrm{critical\ points}:i} (-1)^{d-m_i} \delta^d(\bm{x}-\bm{x}^{(i)}_\mathrm{c})
  = (-1)^d \left(\frac{\sigma_2}{\sigma_1}\right)^d
  \delta^d(\bm{\eta}) \det\zeta,
\end{equation}
where $m_i$ is the index of the $i$th critical point
$\bm{x}^{(i)}_\mathrm{c}$. Due to Eq.~(\ref{eq:1-50}), the expectation
value of the preceding equation with constraint
$\alpha(\bm{x}_\mathrm{c}) \geq \nu$ corresponds to the density of the
Euler characteristic of the body $\mathcal{F}_\nu$:
\begin{equation}
  \label{eq:1-55}
  n_\chi^{(d)}(\nu)
  = (-1)^d\left(\frac{\sigma_2}{\sigma_1}\right)^d
  \left\langle
    \Theta(\alpha - \nu) \delta^d(\bm{\eta}) \det\zeta
  \right\rangle,
\end{equation}
where $\Theta(x)$ is the Heaviside step function.

\section{\label{sec:TheFormula}
  Non-Gaussian corrections to the Minkowski functionals
}

\subsection{\label{subsec:GaussianAvr} Gaussian averages for
  derivatives of the Euler characteristic }

The density of the Euler characteristic $n_\chi^{(d)}(\nu)$ of
$\mathcal{F}_\nu$ in the $d$-dimensional space is given by
Eq.~(\ref{eq:1-55}). Instead of the Euler characteristic, we take the
function $F$ in Sec.~\ref{sec:General} as a differential Euler
characteristic density $-(d/d\nu)n_\chi(\nu)$, i.e.,
\begin{equation}
  \label{eq:2-2}
  F \equiv
  (-1)^d\left(\frac{\sigma_2}{\sigma_1}\right)^d
  \delta(\alpha - \nu) \delta^d(\bm{\eta}) \det\zeta,
\end{equation}
so that the integral of the expectation value $\langle F\rangle$ by
$\nu$ should give the Euler characteristic density $n_\chi(\nu)$:
\begin{equation}
  \label{eq:2-3}
  n_\chi(\nu) = \int_\nu^\infty d\nu \langle F \rangle.
\end{equation}

In order to evaluate Eq.~(\ref{eq:1-15}), we only need to calculate a
Gaussian average of the function $F$ and its derivatives with respect
to $X_\mu$. The Gaussian statistics are characterized only by the
covariance matrix $M_{\mu\nu} = \langle X_\mu X_\nu \rangle$. Assuming
statistical isotropy and rotational invariance of the field variables,
they are given by \cite{BBKS}
\begin{align}
  &
    \left\langle \alpha^2 \right\rangle = 1, \quad
    \left\langle \alpha \eta_i \right\rangle = 0, \quad
    \left\langle \alpha \zeta_{ij} \right\rangle = - \frac{\gamma}{d}
    \delta_{ij}, \quad
    \left\langle \eta_i \eta_j \right\rangle = \frac{1}{d}
    \delta_{ij},
    \nonumber\\
  &
    \left\langle \eta_i \zeta_{jk} \right\rangle = 0, \quad
    \left\langle \zeta_{ij} \zeta_{kl} \right\rangle =
    \frac{1}{d(d+2)}
    \left(\delta_{ij}\delta_{kl} + \delta_{ik}\delta_{jl} +
        \delta_{il}\delta_{jk}  \right),
    \label{eq:2-4}
\end{align}
where
\begin{equation}
    \label{eq:2-5}
    \gamma \equiv \frac{{\sigma_1}^2}{\sigma_0 \sigma_2}.
\end{equation}
It is convenient to define a new variable
\begin{equation}
    \label{eq:2-6}
    Z_{ij} \equiv \frac{d}{\gamma} \zeta_{ij} + \delta_{ij} \alpha,
\end{equation}
instead of $\zeta_{ij}$. The covariances of a new set of variables
$(\alpha,\eta_i,Z_{ij})$ are given by
\begin{align}
  &
    \left\langle \alpha^2 \right\rangle = 1, \quad
    \left\langle \alpha \eta_i \right\rangle = 0, \quad
    \left\langle \alpha Z_{ij} \right\rangle = 0, \quad
  \nonumber\\
  &
    \left\langle \eta_i \eta_j \right\rangle = - \frac{1}{d}
    \delta_{ij}, \quad
    \left\langle \eta_i Z_{jk} \right\rangle = 0,
  \nonumber\\
  &
    \left\langle Z_{ij} Z_{kl} \right\rangle =
    - \delta_{ij}\delta_{kl} +
    \frac{d}{(d+2)\gamma^2}
    \left(
    \delta_{ij}\delta_{kl} + \delta_{ik}\delta_{jl} + \delta_{il}\delta_{jk}
    \right),
  \label{eq:2-7}
\end{align}
so that the variables $\alpha$, $\eta_i$, and $Z_{ij}$ are independent
of each other for Gaussian statistics. In Eq.~(\ref{eq:1-15}), we need
to evaluate the Gaussian average of a type
\begin{align}
  &
  \left\langle
    \frac{\partial^{m_0+2m_1+m_2}F}
    {\partial\alpha^{m_0}
      \partial\eta_{i_1}\cdots\partial\eta_{i_{2m_1}}
      \partial \zeta_{j_1k_1}\cdots\partial\zeta_{j_{m_2}k_{m_2}}}
  \right\rangle_\mathrm{G}
    \nonumber\\
  & \quad
  = \left(\frac{\sigma_1}{d\,\sigma_0}\right)^d
  \left\langle
    \frac{\partial^{2m_1}\delta^d(\bm{\eta})}
    {\partial\eta_{i_1}\cdots\partial\eta_{i_{2m_1}}}
  \right\rangle_\mathrm{G}
  \left(\frac{\gamma}{d}\right)^{-m_2}
    \nonumber\\
  & \qquad
    \times
  \left(-\frac{d}{d\nu}\right)^{m_0}
  \left[
    \left\langle
      \delta(\alpha - \nu)
    \right\rangle_\mathrm{G}
    \left\langle
      \frac{\partial^{m_2}\det\left(\nu I - Z\right)}
      {\partial Z_{j_1k_1}\cdots\partial Z_{j_{m_2}k_{m_2}}}
    \right\rangle_\mathrm{G}
  \right],
  \label{eq:2-8}
\end{align}
where $I$ is a $d\times d$ unit matrix and $Z$ is a $d\times d$
symmetric matrix with $Z_{ji} = Z_{ij}$. Since the Gaussian
distribution functions of the variables $\alpha$ and $\eta_i$ are
given by
\begin{equation}
  \label{eq:2-9}
  P^{(0)}_\mathrm{G}(\alpha) =
  \frac{e^{-\alpha^2/2}}{\sqrt{2\pi}}, \quad
  P^{(1)}_\mathrm{G}(\bm{\eta}) =
  \left(\frac{d}{2\pi}\right)^{d/2}e^{-d|\bm{\eta}|^2/2},
\end{equation}
we have
\begin{align}
  \label{eq:2-10a}
  \left\langle
    \delta(\alpha - \nu)
  \right\rangle_\mathrm{G}
  &= \frac{e^{-\nu^2/2}}{\sqrt{2\pi}},
  \\
  \label{eq:2-10b}
  \left\langle
    \frac{\partial^{2m_1}\delta^d(\bm{\eta})}
    {\partial\eta_{i_1}\cdots\partial\eta_{i_{2m_1}}}
  \right\rangle_\mathrm{G}
  &= 
  \frac{d^{d/2+m_1}}{(2\pi)^{d/2}}
  H_{i_1\cdots i_{2m_1}}(\bm{0}),
\end{align}
where
\begin{equation}
  \label{eq:2-11}
  H_{i_1\cdots i_m}(\bm{x}) = e^{|\bm{x}|^2/2}
  \frac{\partial^m}{\partial x_{i_1}\cdots\partial x_{i_m}}
  e^{-|\bm{x}|^2/2}
\end{equation}
is the multivariate Hermite polynomial. In particular,
\begin{equation}
  \label{eq:2-12}
  H_{i_1\cdots i_{2m}}(\bm{0}) =
    H_{2m}(0) \delta_{(i_1i_2}\cdots\delta_{i_{2m-1}i_{2m})},
\end{equation}
where $H_{2m}(0) = (-1)^m(2m-1)!!$ is the zero-point value of the
(probabilists') Hermite polynomial,
$H_m(\nu) = e^{\nu^2/2}(-d/d\nu)^m e^{-\nu^2/2}$, and the round
brackets in the indices of the Kronecker delta represent
symmetrization of the indices inside the brackets.

\subsection{\label{subsec:UsefulFormulas} A useful formula of Gaussian
  averages for derivatives of the determinant }

Now, we consider the last factor of Eq.~(\ref{eq:2-8}):
\begin{equation}
  \label{eq:2-20}
  \left\langle
    \frac{\partial^{m_2}\det (\nu I - Z)}
    {\partial Z_{j_1k_1}\cdots\partial Z_{j_{m_2}k_{m_2}}}
  \right\rangle_\mathrm{G},
\end{equation}
For $m_2=0$, there is a simple identity
\begin{equation}
  \label{eq:2-21}
  \left\langle \det (\nu I - Z) \right\rangle_\mathrm{G} =
  H_d(\nu).
\end{equation}
The proof of this equation is given in Appendix~\ref{app:LemmaA}.

For $m_2 \geq 1$, the partial derivatives $\partial/\partial Z_{ij}$
are performed under the condition that the variables with $i\leq j$
are the set of independent variables in Eq.~(\ref{eq:2-8}). It is
convenient to introduce a redundant set of independent variables
\begin{equation}
  \label{eq:2-22}
  Y_{ij} \equiv
  \begin{cases}
    Z_{ij}, & (i\leq j), \\
    Z_{ji}, & (i > j),
  \end{cases}
\end{equation}
which is a symmetric tensor, $Y_{ij} = Y_{ji}$. Considering the
variables $Y_{ij}$ as independent variables, the partial derivatives
with respect to $Z_{ij}$ are given by the partial derivatives with
respect to $Y_{ij}$ as
\begin{equation}
  \label{eq:2-23}
  \frac{\partial}{\partial Z_{ij}} =
  \begin{cases}
    \cfrac{\partial}{\partial Y_{ii}}, & (i=j),
    \vspace{1ex} \\
    \cfrac{\partial}{\partial Y_{ij}} + 
    \cfrac{\partial}{\partial Y_{ji}}, & (i<j).
  \end{cases}
\end{equation}
In Eq.~(\ref{eq:1-15}), a type of differential operator,
\begin{equation}
  \label{eq:2-24}
  \sum_{i\leq j} C_{ij} \frac{\partial}{\partial Z_{ij}}
  = \sum_{i,j} C_{ij} \frac{\partial}{\partial Y_{ij}}
  = C_{ij} D_Z^{ij},
\end{equation}
appears, where $C_{ij}$ is an arbitrary symmetric tensor with
$C_{ij} = C_{ji}$, and
\begin{equation}
  \label{eq:2-25}
  D_Z^{ij} \equiv \frac{1}{2}
  \left(
    \frac{\partial}{\partial Y_{ij}} + 
    \frac{\partial}{\partial Y_{ji}}
  \right) =
  \begin{cases}
    \cfrac{\partial}{\partial Z_{ii}}, & (i=j),
    \vspace{1ex} \\
    \cfrac{1}{2}\cfrac{\partial}{\partial Z_{ij}}, & (i<j),
    \vspace{1ex} \\
    \cfrac{1}{2}\cfrac{\partial}{\partial Z_{ji}}, & (i>j)
  \end{cases}
\end{equation}
is a symmetric differential operator, and the summation over repeated
indices is assumed in the last expression of Eq.~(\ref{eq:2-24}).

Thus, when Eq.~(\ref{eq:2-8}) is summed with the weight of the
cumulants, we generally have
\begin{multline}
  \sum_{i_1,\ldots,i_{2m_1}}
  \sum_{j_1 \leq k_1}\cdots\sum_{j_{m_2} \leq k_{m_2}}
  \left\langle
    \alpha^{m_0}
    \eta_{i_1}\cdots\eta_{i_{2m_1}}
    \zeta_{j_1k_1}\cdots\zeta_{j_{m_2}k_{m_2}}
  \right\rangle_\mathrm{c}
  \\
  \times
  \left\langle
    \frac{\partial^{m_0+2m_1+m_2}F}
    {\partial\alpha^{m_0}
      \partial\eta_{i_1}\cdots\partial\eta_{i_{2m_1}}
      \partial \zeta_{j_1k_1}\cdots\partial\zeta_{j_{m_2}k_{m_2}}}
  \right\rangle_\mathrm{G}
  \\
  = \frac{1}{(2\pi)^{(d+1)/2}}
  \left(\frac{\sigma_1}{\sqrt{d}\sigma_0}\right)^d
  d^{m_1} H_{m_1}(0)
  \left(\frac{\gamma}{d}\right)^{-m_2}
  \\
  \times
  \left\langle
    \alpha^{m_0}
    |\bm{\eta}|^{2m_1}
    \zeta_{j_1k_1}\cdots\zeta_{j_{m_2}k_{m_2}}
  \right\rangle_\mathrm{c}
  \left(-\frac{d}{d\nu}\right)^{m_0}
  \\
  \times
  \left[
    e^{-\nu^2/2}  
    \left\langle
      D_Z^{j_1k_1}\cdots D_Z^{j_{m_2}k_{m_2}}\det (\nu I - Z)
    \right\rangle_\mathrm{G}
  \right],
  \label{eq:2-26}
\end{multline}
where both the indices of $\zeta_{jk}$ on the rhs are summed over
$j_i,k_i=1,\ldots,d$ ({\em without} constraints $j_i \leq k_i$).

The partial derivatives $\partial/\partial Z_{ij}$ always appear with
cumulants $C^{(m)}_{\mu_1\cdots\mu_m}$ in Eq.~(\ref{eq:1-15}). These
cumulants are tensors that consist of the Kronecker delta with spatial
indices. Applying the property of Eq.~(\ref{eq:2-24}), we only need a
form
$\langle \mathrm{Tr} ({D_Z}^{m_1}) \cdots \mathrm{Tr} ({D_Z}^{m_k})
\det A\rangle_\mathrm{G}$ to evaluate Eq.~(\ref{eq:1-15}), where
$m_1,\ldots,m_k$, and $k$ are non-negative integers. There is a
remarkable identity,
\begin{multline}
  \left\langle
    \mathrm{Tr} \left({D_Z}^{m_1}\right) \cdots
    \mathrm{Tr}\left({D_Z}^{m_k}\right) \det (\nu I - Z)
  \right\rangle_\mathrm{G}
  \\
  =
  \frac{(-1)^k}{2^{m-k}}
  \frac{d!}{(d-m)!} H_{d-m}(\nu),  
  \label{eq:2-27}
\end{multline}
where $m = m_1 + \cdots + m_k$. Eq.~(\ref{eq:2-21}) is a special case
of $m=k=0$ of this identity. The proof of Eq.~(\ref{eq:2-27}) is given
in Appendix~\ref{app:LemmaB}. This formula plays a central role in
this paper.

\subsection{\label{subsec:Evaluations}
 Evaluations of each order
}

\subsubsection{The Gaussian term}

The first term in the rhs of Eq.~(\ref{eq:1-15}) corresponds to the
Gaussian contribution and is immediately calculated by applying
Eq.~(\ref{eq:2-21}). Putting $m_0=m_1=m_2=0$ in Eq.~(\ref{eq:2-26}),
this term is given by
\begin{equation}
  \label{eq:3-1}
  \langle F \rangle_\mathrm{G} =
  \frac{1}{(2\pi)^{(d+1)/2}}
  \left(\frac{\sigma_1}{\sqrt{d}\sigma_0}\right)^d
  e^{-\nu^2/2}H_d(\nu).
\end{equation}

\subsubsection{The first-order term}

The second term in the rhs of Eq.~(\ref{eq:1-15}) corresponds to the
first-order contribution. We should evaluate 
\begin{multline}
  \label{eq:3-20}
  W^{(1)}
  \equiv
  C^{(3)}_{\mu_1\mu_2\mu_3}
  \left\langle
    \frac{\partial^3F}
    {\partial X_{\mu_1}\partial X_{\mu_2}\partial X_{\mu_3}}
  \right\rangle_\mathrm{G}\sigma_0
  \\
  =
  \left\langle\alpha^3\right\rangle_\mathrm{c}
  \left\langle
    \frac{\partial^3F}{\partial\alpha^3}
  \right\rangle_\mathrm{G} +
  3 \sum_{i\leq j}
  \left\langle\alpha^2\zeta_{ij}\right\rangle_\mathrm{c}
  \left\langle
    \frac{\partial^3F}{\partial\alpha^2\partial\zeta_{ij}}
  \right\rangle_\mathrm{G}
  \\
  +
  3 \sum_{i,j}
  \left\langle\alpha\eta_i\eta_j\right\rangle_\mathrm{c}
  \left\langle
    \frac{\partial^3F}{\partial\alpha\partial\eta_i\partial\eta_j}
  \right\rangle_\mathrm{G}
  \\
  +
  3 \sum_{i\leq j}\sum_{k\leq l}
  \left\langle\alpha\zeta_{ij}\zeta_{kl}\right\rangle_\mathrm{c}
  \left\langle
    \frac{\partial^3F}
    {\partial\alpha\partial\zeta_{ij}\partial\zeta_{kl}}
  \right\rangle_\mathrm{G}
  \\
  +
  3 \sum_{i,j}\sum_{k\leq l}
  \left\langle\eta_i\eta_j\zeta_{kl}\right\rangle_\mathrm{c}
  \left\langle
    \frac{\partial^3F}
    {\partial\eta_i\partial\eta_j\partial\zeta_{kl}}
  \right\rangle_\mathrm{G}
  \\
  +
  \sum_{i\leq j} \sum_{k\leq l} \sum_{p\leq q}
  \left\langle\zeta_{ij}\zeta_{kl}\zeta_{pq}\right\rangle_\mathrm{c}
  \left\langle
    \frac{\partial^3F}
    {\partial\zeta_{ij}\zeta_{kl}\partial\zeta_{pq}}
  \right\rangle_\mathrm{G}.
\end{multline}
The cumulants with odd numbers of spatial indices are zero for
rotational symmetry. The cumulants are independent of the spatial
position, thus we have
\begin{equation}
  \label{eq:3-21}
  \left\langle \alpha^2 \zeta_{ij} \right\rangle_\mathrm{c} =
  \frac{\sigma_0}{\sigma_2}
  \left\langle \alpha^2 \alpha_{,ij} \right\rangle_\mathrm{c} =
  -2\frac{\sigma_0}{\sigma_2}
  \left\langle \alpha \alpha_{,i} \alpha_{,j} \right\rangle_\mathrm{c} =
  - 2\gamma\left\langle \alpha \eta_i\eta_j \right\rangle_\mathrm{c}.
\end{equation}
in the second term of the rhs. Due to Eq.~(\ref{eq:2-26}), the
summation in the fourth term reduces to
\begin{multline}
  \label{eq:3-22}
  \sum_{i\leq j}\sum_{k\leq l}
  \left\langle\alpha\zeta_{ij}\zeta_{kl}\right\rangle_\mathrm{c}
  \left\langle
    \frac{\partial^3F}
    {\partial\alpha\partial\zeta_{ij}\partial\zeta_{kl}}
  \right\rangle_\mathrm{G}
  \\
  =
  \frac{1}{(2\pi)^{(d+1)/2}}
  \left(\frac{\sigma_1}{\sqrt{d}\sigma_0}\right)^d
  \left(\frac{\gamma}{d}\right)^{-2}
  \left\langle
    \alpha
    \zeta_{ij}\zeta_{kl}
  \right\rangle_\mathrm{c}
  \\
  \times
  \left(-\frac{d}{d\nu}\right)
  \left[
    e^{-\nu^2/2}  
    \left\langle
      D_Z^{ij} D_Z^{kl}\det (\nu I - Z)
    \right\rangle_\mathrm{G}
  \right].
\end{multline}
The cumulant in this equation reduces to
\begin{equation}
  \label{eq:3-23}
  \left\langle
    \alpha
    \zeta_{ij}\zeta_{kl}
  \right\rangle_\mathrm{c} =
  - \gamma
  \left\langle
    \eta_i \eta_j \zeta_{kl}
  \right\rangle_\mathrm{c}  -
  \frac{\sigma_0\sigma_1}{{\sigma_2}^2}
  \left\langle
    \alpha \eta_i \alpha_{,jkl}
  \right\rangle_\mathrm{c}.
\end{equation}
The last term on the rhs of this equation does not contribute in
Eq.~(\ref{eq:3-22}). This property is seen as follows: due to the
relation of Eq.~(\ref{eq:b-3}),
$\langle D_Z^{ij}D_Z^{kl}\det A \rangle$ can be replaced by
$\epsilon_{iki_3\cdots i_d}\epsilon_{jlj_3\cdots j_d} \langle
A_{i_3j_3}\cdots A_{i_dj_d}\rangle_\mathrm{G}/(d-2)!$ in
Eq.~(\ref{eq:3-22}). Since the last factor is anti-symmetric with
respect to $(j,l)$, while $\alpha_{,jkl}$ is symmetric, the last term
of Eq.~(\ref{eq:3-23}) does vanish in Eq.~(\ref{eq:3-22}), and only
the first term of Eq.~(\ref{eq:3-23}) survives. In the same manner, it
is observed that the last term of Eq.~(\ref{eq:3-20}) does not
contribute. In fact, the cumulant of the corresponding term is
given by
\begin{equation}
  \label{eq:3-24}
  \left\langle
    \zeta_{ij}\zeta_{kl}\zeta_{pq}
  \right\rangle_\mathrm{c} =
  - \frac{{\sigma_0}^3}{{\sigma_2}^3}
  \left(
    \left\langle
      \alpha_{,i}\alpha_{,jkl}\alpha_{,pq}
    \right\rangle_\mathrm{c} +
    \left\langle
      \alpha_{,i}\alpha_{kl}\alpha_{,jpq}
    \right\rangle_\mathrm{c}
  \right).
\end{equation}
Because the terms on the rhs are symmetric with respect to $(j,l)$ or
$(j,q)$, all the terms of Eq.~(\ref{eq:3-24}) vanish when they are
substituted in the last term of Eq.~(\ref{eq:3-20}).

Therefore, the only cumulants we need to evaluate are
$\langle\alpha^3\rangle_\mathrm{c}$,
$\langle\alpha\eta_i\eta_j\rangle_\mathrm{c}$,
$\langle\eta_i\eta_j\zeta_{kl}\rangle_\mathrm{c}$. For rotational
symmetry, these cumulants are parameterized as
\begin{align}
  &
    \langle\alpha^3\rangle_\mathrm{c} = \tilde{S}^{(0)} \sigma_0, \quad
  \langle\alpha\eta_i\eta_j\rangle_\mathrm{c} =
  \frac{1}{d}\tilde{S}^{(1)} \delta_{ij}\sigma_0,
  \nonumber\\
  &
    \langle\eta_i\eta_j\zeta_{kl}\rangle_\mathrm{c} =
  \frac{\gamma}{d^2}
  \left[
    \tilde{S}^{(2)}_1 \delta_{ij}\delta_{kl} +
    \tilde{S}^{(2)}_2 \Delta_{ij;kl}
  \right]\sigma_0,
  \label{eq:3-25}
\end{align}
where
\begin{equation}
  \label{eq:3-26}
  \Delta_{ij;kl} \equiv \frac{1}{2}
  \left( \delta_{ik}\delta_{jl} +  \delta_{il}\delta_{jk} \right).
\end{equation}
From the identity
\begin{equation}
  \label{eq:3-27}
  \langle\eta_i\eta_j\zeta_{kl}\rangle_\mathrm{c} +
  \langle\eta_i\eta_l\zeta_{jk}\rangle_\mathrm{c} +
  \langle\eta_j\eta_l\zeta_{ik}\rangle_\mathrm{c} = 0,
\end{equation}
we have a relation
\begin{equation}
  \label{eq:3-28}
  \tilde{S}^{(1)}_1 + \tilde{S}^{(1)}_2 = 0.
\end{equation}

Substituting Eqs.~(\ref{eq:3-21}) and (\ref{eq:3-23})--(\ref{eq:3-25})
into Eq.~(\ref{eq:3-20}), omitting the vanishing terms as indicated
above, and using Eqs.~(\ref{eq:2-26}) and (\ref{eq:2-27}), we finally
derive the expression,
\begin{multline}
  \label{eq:3-29}
  W^{(1)} =
  \frac{1}{(2\pi)^{(d+1)/2}}
  \left(\frac{\sigma_1}{\sqrt{d}\sigma_0}\right)^d
  e^{-\nu^2/2}
  \\
  \times
  \Biggl[
    S^{(0)} H_{d+3}(\nu) +
    2d\,S^{(1)} H_{d+1}(\nu)
    \\
    +d(d-1)S^{(2)}_1 H_{d-1}(\nu)
  \Biggr] \sigma_0,
\end{multline}
where we introduce a set of parameters,
\begin{equation}
  \label{eq:3-30}
  S^{(0)} \equiv \tilde{S}^{(0)}, \quad
  S^{(1)} \equiv \frac{3}{2}\tilde{S}^{(1)}, \quad
  S^{(2)} \equiv -\frac{3}{2}\tilde{S}^{(2)}_1,
\end{equation}
for an aesthetic reason.

\subsubsection{The second-order terms}

The second-order terms of Eq.~(\ref{eq:1-15}) are evaluated
similarly to the first-order term. We define
\begin{align}
  \label{eq:3-50a}
  W^{(2)}_1
  &\equiv
  C^{(4)}_{\mu_1\mu_2\mu_3\mu_4}
  \left\langle
    \frac{\partial^4 F}
    {\partial X_{\mu_1}\partial X_{\mu_2}\partial X_{\mu_3}\partial X_{\mu_4}}
  \right\rangle_\mathrm{G} {\sigma_0}^2,
  \\
  \label{eq:3-50b}
  W^{(2)}_2
  &\equiv
  C^{(3)}_{\mu_1\mu_2\mu_3} C^{(3)}_{\mu_4\mu_5\mu_6}
  \left\langle
    \frac{\partial^6 F}
    {\partial X_{\mu_1}\partial X_{\mu_2}\partial X_{\mu_3}
      \partial X_{\mu_4}\partial X_{\mu_5}\partial X_{\mu_6}}
  \right\rangle_\mathrm{G} {\sigma_0}^2.
\end{align}
The evaluation of these terms is similar to the first-order case, and
is straightforward by applying the same techniques. The four-point
cumulants that are required to evaluate $W^{(2)}_1$ are parameterized
as
\begin{align}
  &
    \langle\alpha^4\rangle_\mathrm{c} = \tilde{K}^{(0)} {\sigma_0}^2, \quad
    \langle\alpha^2\eta_i\eta_j\rangle_\mathrm{c} =
    \frac{1}{d}\tilde{K}^{(1)} \delta_{ij} {\sigma_0}^2,
    \nonumber\\
  &
    \langle\alpha\eta_i\eta_j\zeta_{kl}\rangle_\mathrm{c} =
    \frac{\gamma}{d^2}
    \left[
    \tilde{K}^{(2)}_1 \delta_{ij}\delta_{kl} +
    \tilde{K}^{(2)}_2 \Delta_{ij;kl}
    \right] {\sigma_0}^2,
    \nonumber\\
  &
    \langle\eta_i\eta_j\zeta_{kl}\zeta_{pq}\rangle_\mathrm{c} =
    \frac{\gamma^2}{d^3}
    \Biggl[
    \tilde{K}^{(3)}_1 \delta_{ij}\delta_{kl}\delta_{pq}
    \nonumber\\
  & \hspace{6.5pc}
    +
    \frac{1}{2} \tilde{K}^{(3)}_2
    \left(
    \Delta_{ij;kl}\delta_{pq} + \Delta_{ij;pq}\delta_{kl}
    \right)
    \nonumber\\
  & \hspace{6.5pc}
    +
    \tilde{K}^{(3)}_3 \Delta_{kl;pq}\delta_{ij} +
    \tilde{K}^{(3)}_4 \Delta_{ij;kl;pq}
    \Biggr] {\sigma_0}^2,
  \label{eq:3-51}
\end{align}
where
\begin{multline}
  \label{eq:3-52}
  \Delta_{ij;kl;pq} \equiv
  \frac{1}{8}
  \left(
    \delta_{jk}\delta_{lp}\delta_{qi} +
    \delta_{jk}\delta_{lq}\delta_{pi} +
    \delta_{jl}\delta_{kp}\delta_{qi} +
    \delta_{jl}\delta_{kq}\delta_{pi}
    \right.
    \\
    \left.
    +
    \delta_{ik}\delta_{lp}\delta_{qj} +
    \delta_{ik}\delta_{lq}\delta_{pj} +
    \delta_{il}\delta_{kp}\delta_{qj} +
    \delta_{il}\delta_{kq}\delta_{pj}
  \right).
\end{multline}
Other cumulants are given by the above ones as
\begin{align}
  \langle\alpha^3 \zeta_{ij}\rangle_\mathrm{c}
  &=
    -3\gamma \langle\alpha^2 \eta_i\eta_j\rangle_\mathrm{c},
    \nonumber\\
  \langle\alpha^2 \zeta_{ij} \zeta_{kl}\rangle_\mathrm{c}
  &=
    -2\gamma \langle\alpha \eta_i\eta_j\zeta_{kl} \rangle_\mathrm{c}
    + \cdots,
    \nonumber\\
  \langle\alpha \zeta_{ij} \zeta_{kl} \zeta_{pq}\rangle_\mathrm{c}
  &=
    -\gamma \langle \eta_i\eta_j\zeta_{kl} \zeta_{pq}\rangle_\mathrm{c}
    + \cdots,
    \nonumber\\
  \langle \eta_i \eta_j \eta_k \eta_l \rangle_\mathrm{c}
  &=
    -3 \gamma^{-1}
    \langle\alpha \eta_{(i} \eta_j \zeta_{kl)} \rangle_\mathrm{c},
  \label{eq:3-53}
\end{align}
where $+\cdots$ represent terms that do not contribute in
Eq.~(\ref{eq:2-26}). The cumulant
$\langle\zeta_{ij}\zeta_{kl}\zeta_{pq}\zeta_{rs}\rangle_\mathrm{c}$
does not contribute in Eq.~(\ref{eq:2-26}), because of the same reason
we described around Eq.~(\ref{eq:3-24}). From the identity,
\begin{multline}
  \label{eq:3-54}
  \left\langle \eta_i \eta_j \zeta_{kl} \zeta_{pq}
  \right\rangle_\mathrm{c} + 
  \left\langle \eta_i \eta_k \zeta_{jl} \zeta_{pq}
  \right\rangle_\mathrm{c} + 
  \left\langle \eta_j \eta_k \zeta_{il} \zeta_{pq}
  \right\rangle_\mathrm{c}
  \\
  = 
  \left\langle \eta_i \eta_j \zeta_{kq} \zeta_{lp}
  \right\rangle_\mathrm{c} + 
  \left\langle \eta_i \eta_k \zeta_{jq} \zeta_{lp}
  \right\rangle_\mathrm{c} + 
  \left\langle \eta_j \eta_k \zeta_{iq} \zeta_{lp}
  \right\rangle_\mathrm{c},
\end{multline}
we have a relation,
\begin{equation}
  \label{eq:3-55}
  4\tilde{K}^{(3)}_1 +
  2\tilde{K}^{(3)}_2 -
  2\tilde{K}^{(3)}_3 -
  \tilde{K}^{(3)}_4 = 0
\end{equation}
for $d \geq 2$.

The calculations of Eqs.~(\ref{eq:3-50a}) and (\ref{eq:3-50b}) are
similar to the first-order case and are straightforward using
Eqs.~(\ref{eq:2-26}) and (\ref{eq:2-27}) but tedius. The results are
\begin{multline}
  \label{eq:3-56}
  W^{(2)}_1 =
  \frac{1}{(2\pi)^{(d+1)/2}}
  \left(\frac{\sigma_1}{\sqrt{d}\sigma_0}\right)^d
  e^{-\nu^2/2}
  \Bigl\{
    K^{(0)} H_{d+4}(\nu)
    \\
    +
    3d\,K^{(1)} H_{d+1}(\nu) +
    \frac{3d}{2}\left[(d-2)K^{(2)}_1 + d\,K^{(2)}_2\right]
    H_{d-1}(\nu)
    \\
    +\,
    d(d-1)(d-2) K^{(3)} H_{d-2}(\nu)
  \Bigr\} {\sigma_0}^2,
\end{multline}
and
\begin{multline}
  W^{(2)}_2 = 
  \frac{1}{(2\pi)^{(d+1)/2}}
  \left(\frac{\sigma_1}{\sqrt{d}\sigma_0}\right)^d
  e^{-\nu^2/2}
  \Biggl\{
  \left(S^{(0)}\right)^2 H_{d+6}(\nu)
  \\
  +
  4d\,S^{(0)}S^{(1)} H_{d+4}(\nu) +
  2d(d-1)S^{(0)}S^{(2)} H_{d+2}(\nu)
  \\
  +
  4d(d-2)\left(S^{(1)}\right)^2 H_{d+2}(\nu) +
  4d(d-1)(d-4)S^{(1)}S^{(2)} H_{d}(\nu)
  \\
  +
  d(d-1)(d-2)(d-7)\left(S^{(2)}\right)^2
  H_{d-2}(\nu)
  \Biggr\} {\sigma_0}^2,
  \label{eq:3-57}
\end{multline}
where we introduce a set of new parameters,
\begin{align}
  \label{eq:3-57-1}
  &
  K^{(0)} \equiv \tilde{K}^{(0)}, \quad
  K^{(1)} \equiv 2 \tilde{K}^{(1)}, \quad
    K^{(2)}_1 \equiv -2 \tilde{K}^{(2)}_1,
    \nonumber\\
  &
  K^{(2)}_2 \equiv 2 \tilde{K}^{(2)}_2, \quad
  K^{(3)} \equiv 2\tilde{K}^{(3)}_1 - \tilde{K}^{(3)}_3,
\end{align}
for an aesthetic reason.

Eq.~(\ref{eq:1-15}) is given by
\begin{equation}
  \label{eq:3-58}
  \langle F \rangle =
  \langle F \rangle_\mathrm{G} +
  \frac{1}{6} W^{(1)} + \frac{1}{24} W^{(2)}_1 + \frac{1}{72} W^{(2)}_2.
\end{equation}
Substituting Eqs.~(\ref{eq:3-29}), (\ref{eq:3-56}) and (\ref{eq:3-57}),
into Eq.~(\ref{eq:3-58}), and using the integral
\begin{equation}
  \label{eq:3-59}
  \int_\nu^\infty d\nu\,e^{-\nu^2/2}H_n(\nu) = e^{-\nu^2/2}H_{n+1}(\nu),
\end{equation}
one obtains the result for the expression of $n_\chi(\nu)$ of
Eq.~(\ref{eq:2-3}).

\subsection{Skewness and kurtosis parameters}

The skewness and kurtosis parameters can be explicitly given in the
form of rotationally invariant averages of field variables. By taking
all the possible contractions of spatial indices in
Eq.~(\ref{eq:3-25}) and (\ref{eq:3-51}) and solving the resulting
linear equations, one obtains
\begin{align}
  &
  S^{(0)} =
  \frac{\left\langle f^3 \right\rangle_\mathrm{c}}{{\sigma_0}^4},
  \quad
  S^{(1)} = \frac{3}{2}
  \frac{\left\langle f |\bm{\nabla} f|^2\right\rangle_\mathrm{c}}
    {{\sigma_0}^2{\sigma_1}^2},
    \nonumber \\
  &
  S^{(2)} = 
  \frac{-3d}{2(d-1)}
  \frac{\left\langle |\bm{\nabla}f|^2 \Laplace
      f\right\rangle_\mathrm{c}} 
  {{\sigma_1}^4},
  \label{eq:3-70}
\end{align}
and
\begin{align}
  K^{(0)}
  &=
  \frac{\left\langle f^4 \right\rangle_\mathrm{c}}{{\sigma_0}^6},
  \quad
  K^{(1)} =
  2\frac{\left\langle f^2 |\bm{\nabla} f|^2\right\rangle_\mathrm{c}}
  {{\sigma_0}^4{\sigma_1}^2},
  \nonumber\\
  K^{(2)}_1
  &=
  \frac{-2d}{(d+2)(d-1)}
  \frac{(d+2)
    \left\langle f |\bm{\nabla}f|^2 \Laplace f\right\rangle_\mathrm{c} +
    \left\langle |\bm{\nabla}f|^4 \right\rangle_\mathrm{c}} 
  {{\sigma_0}^2{\sigma_1}^4},
  \nonumber\\
  K^{(2)}_2
  &=
  \frac{-2d}{(d+2)(d-1)}
  \frac{(d+2)
    \left\langle
      f |\bm{\nabla}f|^2 \Laplace f
    \right\rangle_\mathrm{c} +
    d \left\langle |\bm{\nabla}f|^4 \right\rangle_\mathrm{c}} 
  {{\sigma_0}^2{\sigma_1}^4},
  \nonumber\\
  K^{(3)}
  &=
  \frac{2d^2}{(d-1)(d-2)}
  \frac{
    \left\langle
    |\bm{\nabla}f|^2 (\Laplace f)^2
    \right\rangle_\mathrm{c} -
    \left\langle
    |\bm{\nabla}f|^2 f_{ij}f_{ij}
    \right\rangle_\mathrm{c}} 
  {{\sigma_1}^6}.
  \label{eq:3-71}
\end{align}
The parameters $K^{(2)}_1$ and $K^{(2)}_2$ are undetermined in the
case of $d=1$, since only the combination $K^{(2)}_1 - K^{(2)}_2$ can
be determined. In this case, by noticing that
$\langle {f_1}^2\rangle = - 3 \langle f {f_1}^2 f_{11}\rangle$, we
have
\begin{equation}
  \label{eq:3-72}
  K^{(2)}_1 - K^{(2)}_2 =
  \frac{2}{3}
  \frac{\langle {f_1}^4 \rangle_\mathrm{c}}{{\sigma_0}^2{\sigma_1}^4},
  \qquad (d=1),
\end{equation}
which is substituted into Eq.~(\ref{eq:3-56}) in the case of $d=1$.
The parameter $K^{(3)}$ is undetermined in the cases of $d=1,2$, but
does not appear in Eq.~(\ref{eq:3-56}). We can ignore the term of
$K^{(3)}$ in these cases.

In Eq.~(\ref{eq:3-70}), the third-order cumulants are the same as the
third-order mean values,
$\langle f^3 \rangle_\mathrm{c} = \langle f^3 \rangle$,
$\langle f |\bm{\nabla}f|^2 \rangle_\mathrm{c} = \langle f
|\bm{\nabla}f|^2 \rangle$,
$\langle |\bm{\nabla}f|^2 \triangle f\rangle_\mathrm{c} = \langle
|\bm{\nabla}f|^2 \triangle f \rangle$ because the mean values are
zero,
$\langle f \rangle = \langle f_i \rangle = \langle f_{ij}\rangle = 0$.
In Eq.~(\ref{eq:3-71}), the fourth-order cumulants are related to the
fourth-order mean values by
\begin{align}
  \label{eq:3-73a}
  \left\langle f^4 \right\rangle_\mathrm{c} &=
  \left\langle f^4 \right\rangle - 3 {\sigma_0}^2,
  \\
  \label{eq:3-73b}
  \left\langle f^2 |\bm{\nabla} f|^2 \right\rangle_\mathrm{c} &=
  \left\langle f^2 |\bm{\nabla} f|^2 \right\rangle
  - {\sigma_0}^2{\sigma_1}^2,
  \\
  \label{eq:3-73c}
  \left\langle f |\bm{\nabla} f|^2 \triangle f \right\rangle_\mathrm{c} &=
  \left\langle f |\bm{\nabla} f|^2 \triangle f \right\rangle
  + {\sigma_1}^4,
  \\
  \label{eq:3-73d}
  \left\langle |\bm{\nabla} f|^4 \right\rangle_\mathrm{c} &=
  \left\langle |\bm{\nabla} f|^4 \right\rangle
  - \frac{d+2}{d} {\sigma_1}^4,
  \\
  \label{eq:3-73e}
  \left\langle |\bm{\nabla} f|^2 (\triangle f)^2 \right\rangle_\mathrm{c} &=
  \left\langle |\bm{\nabla} f|^2 (\triangle f)^2 \right\rangle
  - {\sigma_1}^2{\sigma_2}^2,
  \\
  \label{eq:3-73f}
  \left\langle |\bm{\nabla} f|^2 f_{ij} f_{ij} \right\rangle_\mathrm{c} &=
  \left\langle |\bm{\nabla} f|^2 f_{ij} f_{ij} \right\rangle
  - {\sigma_1}^2{\sigma_2}^2,
\end{align}
which are followed by the definition of cumulants and the fact that
mean values with an odd number of spatial derivatives vanish because
of rotational symmetry.

\subsection{Minkowski functionals}

From Crofton's formula, Eq.~(\ref{eq:1-37}), the $k$th Minkowski
functional in $d$-dimensions, expectation value of the Minkowski
functionals are given by Eq.~(\ref{eq:1-39}), and
$\langle V^{(k)}_k \rangle$ is the expectation value of the density of
the Euler characteristic. Because the formula for the Euler
characteristic we have derived thus far is for general $d$ dimensions,
the last quantity $\langle V^{(k)}_k \rangle$ can simply replace
$n_\chi(\nu)$ with $d=k$, assuming that all the parameters are
calculated in $k$-dimensional subspace. As such, the parameters in the
derived formula,
$\sigma_0, \sigma_1, S^{(0)}, \ldots, K^{(0)}, \ldots$, should be
replaced by the corresponding parameters in the $k$-dimensional
subspace in $d$-dimensional space.

We denote the corresponding parameters as
${}^k\!\sigma_0, {}^k\!\sigma_1, {}^k\!S^{(0)}, \ldots,
{}^k\!K^{(0)}, \ldots$. These parameters are represented by
corresponding ones in $d$-dimensional space as
\begin{align}
  \label{eq:4-2a}
  {{}^k\!\sigma_0}^2
  &= \langle f^2\rangle = {\sigma_0}^2,
  \\
  \label{eq:4-2b}
  {{}^k\!\sigma_1}^2
  &= -\langle f\Laplace_k f\rangle
  = -k\langle f f_{11}\rangle
  = -\frac{k}{d}\langle f \Laplace f\rangle
  = \frac{k}{d} {\sigma_1}^2,
\end{align}
where $\Laplace_k$ is the Laplacian operator in $k$-dimensional
subspace. Similarly, we have
\begin{align}
  \label{eq:4-3a}
  {}^k\!S^{(0)}
  &=
    \frac{\langle f^3 \rangle_\mathrm{c}}{{{}^k\!\sigma_0}^4}
    = \frac{\langle f^3\rangle_\mathrm{c}}{{\sigma_0}^4}
    = \frac{\langle \alpha^3\rangle_\mathrm{c}}{\sigma_0}
    = S^{(0)},
  \\
  \label{eq:4-3b}
  {}^k\!S^{(1)}
  &= \frac{3}{2}
    \frac{\langle f|\bm{\nabla}_{\!k}f|^2 \rangle_\mathrm{c}}
    {{{}^k\!\sigma_0}^2\ {{}^k\!\sigma_1}^2 }
    = \frac{3}{2} \frac{k \langle f {f_1}^2\rangle_\mathrm{c}}
    {{\sigma_0}^2 (k/d){\sigma_1}^2}
    = \frac{3}{2} \frac{d \langle \alpha {\eta_1}^2\rangle_\mathrm{c}}{\sigma_0}
    = S^{(1)},
  \\
  \label{eq:4-3c}
  {}^k\!S^{(2)}
  &= 
    \frac{-3k}{2(k-1)}
    \frac{\langle
    |\bm{\nabla}_{\!k}f|^2\Laplace_kf
    \rangle_\mathrm{c}}{{{}^k\!\sigma_1}^4}
    = \frac{-3k}{2(k-1)}
    \frac{
    k\langle {f_1}^2 \Laplace_kf\rangle_\mathrm{c}}
    {(k/d)^2{\sigma_1}^4}
    \nonumber\\
  &=
    \frac{-3d^2}{2(k-1)} \frac{{\sigma_1}^2 \sigma_2}{{\sigma_1}^4}
    \left[\langle {\eta_1}^2 \zeta_{11}\rangle_\mathrm{c} +
    (k-1)\langle {\eta_1}^2 \zeta_{22}\rangle_\mathrm{c}\right]
    = S^{(2)},
\end{align}
where $\bm{\nabla}_{\!k}$ is the gradient in the subspace, and
Eq.~(\ref{eq:3-25}) is applied to derive the last expressions. For the
kurtosis parameters, similar calculations show that
\begin{align}
  &
    {}^k\!K^{(0)} = K^{(0)}, \quad
  {}^k\!K^{(1)} = K^{(1)}, \quad
  {}^k\!K^{(2)}_1 = K^{(2)}_1,
    \nonumber\\
  &
  {}^k\!K^{(2)}_2 = K^{(2)}_2, \quad
  {}^k\!K^{(3)} = K^{(3)}.
  \label{eq:4-4}
\end{align}

Combining Eqs.~(\ref{eq:2-3}), (\ref{eq:3-29}),
(\ref{eq:3-56})--(\ref{eq:3-59}) and (\ref{eq:4-2a})--(\ref{eq:4-4}),
a weakly non-Gaussian formula for the density of Minkowski functionals
is finally derived as
\begin{widetext}
\begin{align}
    \bar{V}^{(d)}_k(\nu)
    &=
    \frac{1}{(2\pi)^{(k+1)/2}}
    \frac{\omega_d}{\omega_{d-k}\omega_k}
    \left(\frac{\sigma_1}{\sqrt{d}\sigma_0}\right)^k
    e^{-\nu^2/2}
    \Biggl[\!\Biggl[
    H_{k-1}(\nu) +
    \left[
    \frac{1}{6}S^{(0)} H_{k+2}(\nu) +
    \frac{k}{3} S^{(1)} H_{k}(\nu)
    +
    \frac{k(k-1)}{6} S^{(2)} H_{k-2}(\nu)
    \right] \sigma_0
    \nonumber\\
  & \qquad
    +
    \Biggl\{
    \frac{1}{72} (S^{(0)})^2 H_{k+5}(\nu)
    +
    \left(
    \frac{1}{24}K^{(0)} + \frac{k}{18} S^{(0)}S^{(1)}
    \right) H_{k+3}(\nu)
    + k
    \left[
    \frac{1}{8} K^{(1)} +
    \frac{k-1}{36} S^{(0)} S^{(2)} +
    \frac{k-2}{18} (S^{(1)})^2
    \right] H_{k+1}(\nu)
    \nonumber\\
  & \hspace{5pc}
    + k
    \left[
    \frac{k-2}{16} K^{(2)}_1 + \frac{k}{16} K^{(2)}_2 +
    \frac{(k-1)(k-4)}{18} S^{(1)}S^{(2)}
    \right] H_{k-1}(\nu)
    \nonumber\\
  & \hspace{8pc}
    +
    k(k-1)(k-2)
    \left[
    \frac{1}{24} K^{(3)} + \frac{k-7}{72} (S^{(2)})^2
    \right] H_{k-3}(\nu)
    \Biggr\} {\sigma_0}^2
    + \mathcal{O}\left({\sigma_0}^3\right)
    \Biggr]\!\Biggr].
    \label{eq:4-6}
\end{align}
\end{widetext}
This is the main result of this paper. We confirm that specific cases
of this general result agree with all the known results in the
literature. The lowest-order term, i.e., the Gaussian part agrees with
Tomita's formula \cite{Tom86}. The first-order term
$\mathcal{O}(\sigma_0)$ is in exact agreement with the result of
Ref.~\cite{TM03}, which is a conjectured equation suggested by
lower-dimensional calculations. Therefore, the newly obtained result
is a proof of this conjecture for general dimensions. The second-order
term $\mathcal{O}({\sigma_0}^2)$ in $d=2$ dimensions is in exact
agreement with the result of Ref.~\cite{TM10}. The second-order term
of $V^{(d)}_d(\nu)$, which is equivalent to the genus statistic up to
the overall amplitude, exactly agrees with the results of
Ref.~\cite{Cod13} in $d=2,3$ dimensions, after the conversion of the
cumulants in this reference to the skewness and kurtosis parameters in
this paper.

\subsection{Spectral representation of parameters}

In cosmological applications of Minkowski functionals, it is
convenient to represent the parameters in the derived formula in terms
of the power spectrum $P(k)$, bispectrum
$B(\bm{k}_1,\bm{k}_2,\bm{k}_3)$ and trispectrum
$T(\bm{k}_1,\bm{k}_2,\bm{k}_3,\bm{k}_4)$ of the field $f$, because
these polyspectra can be directly predicted from theories such as the
higher-order perturbation theory of non-linear gravitational
evolution, etc. Although the relations are relatively straightforward,
we explicitly provide the relations in the following for convenience.

Denoting the Fourier transform of the field as
\begin{equation}
    \label{eq:4-20}
    \tilde{f}(\bm{k}) = \int d^d\!x\,e^{-i\bm{k}\cdot\bm{x}} f(\bm{x}),
\end{equation}
the polyspectra up to fourth order are defined as
\begin{align}
  \label{eq:4-21a}
  &
  \left\langle
  \tilde{f}(\bm{k}) \tilde{f}(\bm{k}')
  \right\rangle_\mathrm{c}
  = (2\pi)^d \delta^d(\bm{k}+\bm{k}') P(k),
  \\
  \label{eq:4-21b}
  &
  \left\langle
  \tilde{f}(\bm{k}_1) \tilde{f}(\bm{k}_2) \tilde{f}(\bm{k}_3)
  \right\rangle_\mathrm{c}
  = (2\pi)^d \delta^d(\bm{k}_1+\bm{k}_2+\bm{k}_3)
    B(\bm{k}_1,\bm{k}_2,\bm{k}_3),
  \\
  &
  \left\langle
  \tilde{f}(\bm{k}_1) \tilde{f}(\bm{k}_2) \tilde{f}(\bm{k}_3)
  \tilde{f}(\bm{k}_4) 
    \right\rangle_\mathrm{c}
    \nonumber\\
  \label{eq:4-21c}
  & \qquad
  = (2\pi)^d \delta^d(\bm{k}_1+\bm{k}_2+\bm{k}_3+\bm{k}_4)
    T(\bm{k}_1,\bm{k}_2,\bm{k}_3,\bm{k}_4).
\end{align}
The parameters defined by Eqs.~(\ref{eq:1-11}), (\ref{eq:3-70}) and
(\ref{eq:3-71}) are represented by these spectra as
\begin{align}
  \label{eq:4-22a}
  {\sigma_j}^2
  &= \int \frac{d^dk}{(2\pi)^d} k^{2j} P(k),
  \\
  S^{(a)}
  &= \frac{1}{{\sigma_0}^{4-2a}{\sigma_1}^{2a}}
    \int \frac{d^dk_1}{(2\pi)^d} \frac{d^dk_2}{(2\pi)^d}
    \frac{d^dk_3}{(2\pi)^d}
    \nonumber\\
  \label{eq:4-22b}
  & \quad \times
    (2\pi)^d\delta^d(\bm{k}_1+\bm{k}_2+\bm{k}_3)
    s^{(a)}(\bm{k}_1,\bm{k}_2,\bm{k}_3)
    B(\bm{k}_1,\bm{k}_2,\bm{k}_3),
  \\
  K^{(a)}_\cdot
  &= \frac{1}{{\sigma_0}^{6-2a}{\sigma_1}^{2a}}
    \int \frac{d^dk_1}{(2\pi)^d} \frac{d^dk_2}{(2\pi)^d}
    \frac{d^dk_3}{(2\pi)^d}  \frac{d^dk_4}{(2\pi)^d} 
    \nonumber\\
  & \hspace{4pc} \times
    (2\pi)^d\delta^d(\bm{k}_1+\bm{k}_2+\bm{k}_3+\bm{k}_4)
    \nonumber\\
  \label{eq:4-22c}
  & \hspace{4pc} \times
    \kappa^{(a)}_\cdot(\bm{k}_1,\bm{k}_2,\bm{k}_3,\bm{k}_4)
    T(\bm{k}_1,\bm{k}_2,\bm{k}_3,\bm{k}_4),
\end{align}
where
\begin{align}
  &
    s^{(0)} = 1, \quad
    s^{(1)} = -\frac{3}{2} \bm{k}_1\cdot\bm{k}_2,
    \nonumber \\
  &
    s^{(2)} = -\frac{3d}{2(d-1)} (\bm{k}_1\cdot\bm{k}_2){k_3}^2
    \nonumber \\
  &
    \kappa^{(0)} = 1, \quad
    \kappa^{(1)} = -2\bm{k}_1\cdot\bm{k}_2,
    \nonumber \\
  &
    \kappa^{(2)}_1 =
    \frac{-2d}{(d+2)(d-1)}(\bm{k}_1\cdot\bm{k}_2)
    \left[
    (d+2){k_3}^2 + \bm{k}_3\cdot\bm{k}_4
    \right],
  \nonumber\\
  &
    \kappa^{(2)}_2 =
    \frac{-2d}{(d+2)(d-1)}(\bm{k}_1\cdot\bm{k}_2)
    \left[
    (d+2){k_3}^2 + d\, \bm{k}_3\cdot\bm{k}_4
    \right],
  \nonumber\\
  \label{eq:4-23}
  &
    \kappa^{(3)} =
    \frac{-2d^2}{(d-1)(d-2)}(\bm{k}_1\cdot\bm{k}_2)
    \left[
    {k_3}^2{k_4}^2 - (\bm{k}_3\cdot\bm{k}_4)^2
    \right].
\end{align}

Once the functional forms of the power spectrum, bispectrum and
trispectrum in a model are given, the parameters of the model can be
calculated by the above equations. The dimensionality of the integrals
of Eqs.~(\ref{eq:4-22b}) and (\ref{eq:4-22c}) are too large to
evaluate straightforwardly in higher-dimensional spaces. For the
non-linear perturbation theory of gravitational evolution \cite{Ber02}
in three-dimensional space, one can in principle apply a technique
developed in Refs.~\cite{SVM16,SV16,MFHB16,FBMH17} to reduce the
dimensionality of multi-dimensional integrations of perturbation
kernels. Explicit implementation of the algorithm will be addressed in
a subsequent work \cite{MHK20}.

\section{\label{sec:Conclusions}
  Conclusions
}

In this paper, we present a method to analytically calculate the
non-Gaussian corrections of the Minkowski functionals for the
excursion set of smoothed fields in general dimensions. We explicitly
derive analytic formulas for first- and second-order corrections of
non-Gaussianity for the Minkowski functionals, Eq.~(\ref{eq:4-6}),
which is the main result of the paper. In the derivation, the formula
of Eq.~(\ref{eq:2-27}) plays a central role. It is straightforward to
generalize our calculations to higher-order corrections.

The findings of this paper are quite general. Non-Gaussian corrections
to the expected Minkowski functionals of an excursion set is generally
given in arbitrary dimensions $d$, based on the assumptions of
statistical homogeneity and isotropy of space only. In cosmology, the
cases $d=1,2,3$ are of particular interest for the analyses of cosmic
fields. The formulas for the first-order corrections with $d=1,2,3$,
which were derived in a previous work \cite{TM94,TM03}, are reproduced
from our general formula as special cases. The formulas for
second-order corrections with $d=2$, which have been reported in a
previous work \cite{TM10}, are also reproduced as a special case.
Moreover, the formulas for the second-order corrections for the Euler
characteristic with $d=2,3$, which were derived in a previous work
\cite{Cod13}, are also reproduced as special cases of the general
formula. Thus, our formula contains all the previously known formulas
as special cases, and unifies them into a single formula, generalizing
them to arbitrary dimensions.

The non-Gaussian corrections to the Minkowski functionals are
parameterized by the skewness and kurtosis parameters defined by
Eqs.~(\ref{eq:3-70}) and (\ref{eq:3-71}). In cosmic fields, these
parameters can be theoretically predicted in principle, provided that
the bispectrum and trispectrum are known. The relations are given by
Eqs.~(\ref{eq:4-22a})--(\ref{eq:4-22c}). They involve
multi-dimensional integrations that are not easy to evaluate
numerically in a straightforward manner, especially for the kurtosis
parameters in higher-dimensional space. In practice, one can apply a
technique developed in Refs.~\cite{SVM16,SV16,MFHB16,FBMH17} to reduce
the dimensionality of integration, and the multi-dimensional
integration is reduced to be evaluated by one-dimensional fast Fourier
transforms with \textsc{FFTLog} developed by Hamilton \cite{Ham00}.
Future work should focus on an approach along this line to
theoretically predict the skewness and kurtosis parameters for the
case of cosmic fields, such as the three-dimensional density field,
two-dimensional weak lensing fields, and so on, with bispectra and
trispectra predicted from various theoretical models.

Comparisons of the predicted Minkowski functionals with those
calculated from numerical realizations of weakly non-Gaussian random
fields in two dimensions have already been performed in
Ref.~\cite{TM10} with first- and second-order corrections of the
non-Gaussianity. The results are in complete agreement with each other
within the limit of numerical errors of the realizations. Detailed
comparisons of the first- and second-order corrections with numerical
realizations for three dimensions, and comparisons with data of
cosmological $N$-body simulations will be presented in a subsequent
paper \cite{MHK20}.

Considering the derived formula beyond three dimensions, it would be
extraordinary interesting if the cases of $d\geq 4$ could be applied
to some sort of abstract data analysis, or higher-dimensional theories
in fundamental physics, etc.

\begin{acknowledgments}
  The authors thank S.~Ikeda and T.~T.~Takeuchi for organizing the
  workshop on Minkowski functionals at IPMU (June 20th, 2017), where
  this project was initiated. This work was supported by JSPS KAKENHI
  Grants No.~JP19K03835 (T.M.) and No.~JP16H02792 (S.K.).
\end{acknowledgments}

\newpage

\appendix
% \onecolumngrid

\section{\label{app:LemmaA}
  Proof of Eq.~(\ref{eq:2-21})
}

In this appendix, Eq.~(\ref{eq:2-21}),
\begin{equation}
  \label{eq:a-1}
  \left\langle \det (\nu I - Z) \right\rangle_\mathrm{G} = H_d(\nu).
\end{equation}
is proven. This formula is already known (see, e.g.,
Refs.~\cite{Tom90,Adl07}). Here, we provide an alternative proof. We
define
\begin{equation}
  \label{eq:a-1-1}
  A = \nu I - Z
\end{equation}
below.

Considering the derivative of $\det A$ with respect to $\nu$, we have
\begin{equation}
  \label{eq:a-2}
  \frac{\partial}{\partial\nu} \det A =
  \frac{\partial A_{ij}}{\partial\nu}
  \frac{\partial\det A}{\partial A_{ij}} =
  \sum_{i=1}^d \hat{A}_{i},
\end{equation}
where $ \hat{A}_i$ is the $(i,i)$ minor of the matrix $A$. Given the
statistical isotropy, we have
\begin{equation}
  \label{eq:a-3}
  \left\langle \hat{A}_i \right\rangle_\mathrm{G} =
  \left\langle \det A^{(d-1)} \right\rangle_\mathrm{G},
\end{equation}
where $A^{(d-1)}=\nu I^{(d-1)} - Z^{(d-1)}$ is the matrix $A$
in $(d-1)$-dimensional subspace. Thus, we have
\begin{equation}
  \label{eq:a-4}
  \frac{d}{d\nu}\left\langle \det A \right\rangle_\mathrm{G}  = 
  d\, \left\langle \det A^{(d-1)} \right\rangle_\mathrm{G}.
\end{equation}
Using the same approach, we can show a recursion relation
\begin{equation}
  \label{eq:a-5}
  \frac{d}{d\nu}\left\langle \det A^{(m)} \right\rangle_\mathrm{G}  = 
  m \left\langle \det A^{(m-1)} \right\rangle_\mathrm{G}.
\end{equation}
for $m=1,2,\ldots$, where $\det A^{(0)} \equiv 1$. 
Thereby,
\begin{equation}
  \label{eq:a-6}
  \tilde{H}_m(\nu) \equiv
  \left\langle \det A^{(m)} \right\rangle_\mathrm{G}   
\end{equation}
satisfies the same relation as the Appell
sequence of Hermite polynomials,
\begin{equation}
  \label{eq:a-7}
  H_m'(\nu) = mH_{m-1}(\nu). 
\end{equation}
Therefore, if the integration constant $\tilde{H}_m(0)$ is the same as
$H_m(0)$, $\tilde{H}_m(\nu)$ is identified with $H_m(\nu)$ by
induction. The relation $\tilde{H}_m(0) = H_m(0)$ can be shown as
follows. We have
\begin{equation}
  \label{eq:a-8}
  \tilde{H}_m(0) =
  (-1)^m \left\langle \det Z^{(m)} \right\rangle_\mathrm{G}.
\end{equation}
The determinant of the matrix $Z^{(m)}$ in $m$-dimensional subspace is
given by
\begin{equation}
  \label{eq:a-9}
  \det Z^{(m)} = \frac{1}{m!}
  \epsilon_{i_1\cdots i_m} \epsilon_{j_1\cdots j_m}
  Z^{(m)}_{i_1j_1} \cdots Z^{(m)}_{i_mj_m}.
\end{equation}
The Gaussian average of this equation with odd $m$ is zero. For odd
$m$, we have $H_m(0) = 0$ and Eq.~(\ref{eq:a-8}) is trivially
identified with $H_m(0)$. For even $m$, we have
\begin{multline}
  \label{eq:a-10}
  \left\langle \det Z^{(m)} \right\rangle_\mathrm{G}
  \\
  = \frac{(m-1)!!}{m!} \epsilon_{i_1\cdots i_m}
  \epsilon_{j_1\cdots j_m}
  \left\langle Z^{(m)}_{i_1j_1} Z^{(m)}_{i_2j_2}\right\rangle \cdots
  \left\langle Z^{(m)}_{i_{m-1}j_{m-1}} Z^{(m)}_{i_mj_m}\right\rangle,
\end{multline}
where the Wick's probability theorem for the multivariate normal
distribution is applied. Eq.~(\ref{eq:2-7}) also holds for
$Z^{(m)}$ in the $m$-dimensional subspace with $1\leq i,j,k,l \leq m$,
\begin{equation}
  \label{eq:a-11}
  \left\langle Z^{(m)}_{ij} Z^{(m)}_{kl} \right\rangle =
    - \delta_{ij}\delta_{kl} +
    \frac{d}{(d+2)\gamma^2}
    \left(
    \delta_{ij}\delta_{kl} + \delta_{ik}\delta_{jl} + \delta_{il}\delta_{jk}
    \right),
\end{equation}
In Eq.~(\ref{eq:a-10}), symmetric components with respect to the
permutations of $(i_1,i_2)$, $(j_1,j_2)$, etc., should vanish, and thus
we can substitute
\begin{multline}
  \label{eq:a-12}
  \left\langle Z^{(m)}_{i_1j_1} Z^{(m)}_{i_2j_2} \right\rangle
  \rightarrow
  \frac{1}{4}
  \left[
    2\left\langle Z^{(m)}_{i_1j_1} Z^{(m)}_{i_2j_2} \right\rangle -
      \left\langle Z^{(m)}_{i_2j_1} Z^{(m)}_{i_1j_2} \right\rangle -
      \left\langle Z^{(m)}_{i_1j_2} Z^{(m)}_{i_2j_1} \right\rangle
    \right]
    \\
    =
    - \frac{1}{2}
    \left(
      \delta_{i_1j_1} \delta_{i_2j_2} -
      \delta_{i_1j_2} \delta_{i_2j_1}\right),
\end{multline}
etc.~in Eq.~(\ref{eq:a-10}). Consequently, we have
\begin{equation}
  \label{eq:a-13}
  \left\langle \det Z^{(m)} \right\rangle_\mathrm{G} =
  (m-1)!! (-1)^{m/2} = H_m(0),
\end{equation}
where $\epsilon_{i_1\cdots i_m}\epsilon_{i_1\cdots i_m} = m!$ is used.
Therefore, $\tilde{H}_m(0)$ in Eq.~(\ref{eq:a-8}) is identified with
$H_m(0)$ for even $m$. Thus, we have
$\langle\det A^{(m)}\rangle_\mathrm{G} = H_m(\nu)$. Setting $m=d$ in
this relation completes the proof of Eq.~(\ref{eq:a-1}).

\section{\label{app:LemmaB}
  Proof of Eq.~(\ref{eq:2-27})
}

In this appendix, we prove Eq.~(\ref{eq:2-27}):
\begin{multline}
  \label{eq:b-1}
  \left\langle
    \mathrm{Tr} \left({D_Z}^{m_1}\right) \cdots
    \mathrm{Tr}\left({D_Z}^{m_k}\right) \det (\nu I - Z)
  \right\rangle_\mathrm{G}
  \\
  =
  \frac{(-1)^k}{2^{m-k}}
  \frac{d!}{(d-m)!} H_{d-m}(\nu),
\end{multline}
where $m = m_1 + \cdots + m_k$ and $m_1,\ldots,m_k$ are non-negative
integers, and $D_Z$ is given by Eq.~(\ref{eq:2-25}).

The determinant of $A = \nu I - Z$ is given by
\begin{equation}
  \label{eq:b-2}
  \det A = \frac{1}{n!}
  \epsilon_{i_1\cdots i_d} \epsilon_{j_1\cdots j_d}
  A_{i_1j_1}\cdots A_{i_dj_d}.
\end{equation}
From this expression, we calculate
\begin{align}
  &
    D_Z^{i_1j_1}\cdots D_Z^{i_mj_m} \det A
    \nonumber\\
  &= \frac{(-1)^m}{2^m}
    \left(
    \frac{\partial}{\partial A_{i_1j_1}} +
    \frac{\partial}{\partial A_{j_1i_1}}
    \right) \cdots
    \left(
    \frac{\partial}{\partial A_{i_mj_m}} +
    \frac{\partial}{\partial A_{j_mi_m}}
    \right) \det A
    \nonumber\\
  &= \frac{(-1)^m}{2^m(d-m)!}
    \epsilon_{i_1\cdots i_d} \epsilon_{j_1\cdots j_d}
    A_{i_{m+1}j_{m+1}}\cdots A_{i_dj_d} +
    \mathrm{sym.}
    \left(
    \begin{matrix}
      i_1 \leftrightarrow j_1 \\
      \vdots \\
      i_m \leftrightarrow j_m 
    \end{matrix}
  \right).
  \label{eq:b-3}
\end{align}
Thus we have
\begin{align}
  &
    \mathrm{Tr}\left({D_Z}^m\right) \det A
  =
  D_Z^{k_1k_2}D_Z^{k_2k_3}\cdots D_Z^{k_mk_1} \det A
  \nonumber \\
  &=
  \frac{2(-1)^m}{2^m (d-m)!}
  \epsilon_{k_1k_2\cdots k_mi_{m+1}\cdots i_d}
  \epsilon_{k_2k_3\cdots k_mk_1j_{m+1}\cdots j_d}
  A_{i_{m+1}j_{m+1}}\cdots A_{i_dj_d}
  \nonumber \\
  &=
  \frac{-1}{2^{m-1} (d-m)!}
  \epsilon_{k_1k_2\cdots k_mi_{m+1}\cdots i_d}
  \epsilon_{k_1k_2\cdots k_mj_{m+1}\cdots j_d}
  A_{i_{m+1}j_{m+1}}\cdots A_{i_dj_d},
  \label{eq:b-4}
\end{align}
where the second equality is derived sincethere is no contribution
when the same indices appear in the anti-symmetric tensor.

We also calculate
\begin{multline}
  \label{eq:b-5}
  \frac{\partial^m}{\partial\nu^m} \det A =
  \frac{\partial}{\partial A_{k_1k_1}} \cdots
  \frac{\partial}{\partial A_{k_mk_m}} \det A
  \\
  =
  \frac{1}{(d-m)!}
  \epsilon_{k_1k_2\cdots k_mi_{m+1}\cdots i_d}
  \epsilon_{k_1k_2\cdots k_mj_{m+1}\cdots j_d}
  A_{i_{m+1}j_{m+1}}\cdots A_{i_dj_d}.
\end{multline}
Comparing Eqs.~(\ref{eq:b-4}) and (\ref{eq:b-5}), we have
\begin{equation}
  \label{eq:b-6}
  \mathrm{Tr}\left({D_Z}^m\right) \det A =
  \frac{-1}{2^{m-1}}
  \frac{\partial^m}{\partial\nu^m}\det A,
\end{equation}
and consequently, we have
\begin{equation}
  \label{eq:b-7}
  \mathrm{Tr}\left({D_Z}^{m_1}\right) \cdots
  \mathrm{Tr}\left({D_Z}^{m_k}\right) \det A =
  \frac{(-1)^k}{2^{m-k}}
  \frac{\partial^m}{\partial\nu^m}\det A,
\end{equation}
where $m=m_1+\cdots m_k$. Taking the Gaussian average of the last
equation and applying Eq.~(\ref{eq:a-1}), we have
\begin{equation}
  \label{eq:b-8}
  \left\langle
  \mathrm{Tr}\left({D_Z}^{m_1}\right) \cdots
  \mathrm{Tr}\left({D_Z}^{m_k}\right) \det A
  \right\rangle_\mathrm{G} =
  \frac{(-1)^k}{2^{m-k}}
  \frac{d^m}{d\nu^m} H_d(\nu).
\end{equation}
Eq.~(\ref{eq:b-1}) immediately follows from the repeated use of the
Appell sequence of Eq.~(\ref{eq:a-7}).

% \begin{figure}%[t]
% \begin{center}
% \includegraphics[width=18pc]{FigA01.eps}%{iPTdiag1.eps}
% \caption{\label{fig:iPTdiag1}
% Diagrammatic rules of the iPT: dynamics and biasing.
% }
% \end{center}
% \end{figure}

%%%%%%%%%%%%
\renewcommand{\apj}{Astrophys.~J. }
\newcommand{\aap}{Astron.~Astrophys. }
\newcommand{\aj}{Astron.~J. }
\newcommand{\apjl}{Astrophys.~J.~Lett. }
\newcommand{\apjs}{Astrophys.~J.~Suppl.~Ser. }
\newcommand{\apss}{Astrophys.~Space Sci. }
\newcommand{\cqg}{Class.~Quant.~Grav. }
\newcommand{\jcap}{J.~Cosmol.~Astropart.~Phys. }
\newcommand{\mnras}{Mon.~Not.~R.~Astron.~Soc. }
\newcommand{\mpla}{Mod.~Phys.~Lett.~A }
\newcommand{\pasj}{Publ.~Astron.~Soc.~Japan }
\newcommand{\physrep}{Phys.~Rep. }
\newcommand{\ptp}{Progr.~Theor.~Phys. }
\newcommand{\ptep}{Prog.~Theor.~Exp.~Phys. }
\newcommand{\jetp}{JETP }
\newcommand{\jhep}{Journal of High Energy Physics}
%\newcommand{\prl}{Phys. Rev. Lett.}
%\renewcommand{\prd}{Phys.~Rev.~D}

%\bibliography{redoneloop}% Produces the bibliography via BibTeX.

% \twocolumngrid

\end{document}